\documentclass{emulateapj}
\usepackage{color,verbatim} 

\usepackage{epsfig}
\usepackage{subfigure}
\usepackage{graphicx}
\usepackage{amsmath}
\usepackage{natbib}
\newcommand{\pa}{\partial}
\newcommand{\mb}{\boldsymbol}

\shorttitle{Modeling Emission from Gamma-ray Pulsars using Force-Free Field}
\shortauthors{Bai \& Spitkovsky}

\begin{document}

\title{Modeling of Gamma-Ray Pulsar Light Curves using Force-Free Magnetic Field}

\author{Xue-Ning Bai \& Anatoly Spitkovsky}
\affil{Department of Astrophysical Sciences, Princeton University,
Princeton, NJ, 08544} \email{xbai@astro.princeton.edu,
anatoly@astro.princeton.edu}

\begin{abstract}
Gamma-ray emission from pulsars has long been modeled using a vacuum
dipole field. This approximation ignores changes in the field structure
caused by the magnetospheric plasma and strong plasma currents. We present
the first results of gamma-ray pulsar light curve modeling using the more
realistic field taken from three-dimensional force-free magnetospheric
simulations. Having the geometry of the field, we apply several prescriptions
for the location of the emission zone, comparing the light curves to
observations. We find that when the emission region is chosen according to
the conventional slot-gap (or two-pole caustic) prescription, the model fails
to produce double-peak pulse profiles, mainly because the size of the polar
cap in force-free magnetosphere is larger than the vacuum field polar cap.
This suppresses caustic formation in the inner magnetosphere. The
conventional outer-gap model is capable of producing only one peak under
general conditions, because a large fraction of open field lines does not
cross the null charge surface. We propose a novel ``separatrix layer" model,
where the high-energy emission originates from a thin layer on the open field
lines just inside of the separatrix that bounds the open flux tube. The
emission from this layer generates two strong caustics on the sky map due to
the effect we term ``Sky Map Stagnation" (SMS). It is related to the fact that
the force-free field asymptotically approaches the field of a rotating split
monopole, and the photons emitted on such field lines in the outer magnetosphere
arrive to the observer in phase. The double-peak light curve is a natural
consequence of SMS. We show that most features of the currently available
gamma-ray pulsar light curves can be reasonably well reproduced and explained
with the separatrix layer model using the force-free field. Association of the
emission region with the current sheet will guide more detailed future studies
of the magnetospheric acceleration physics.
\end{abstract}



\section{Introduction}

Gamma-ray pulsars are of great scientific interest in high-energy
astrophysics. The high energy emission from these objects
provides valuable information about pulsar magnetospheric structure,
particle acceleration mechanisms, and plasma physics in strong
magnetic fields. Observations by the Energetic Gamma Ray
Experiment Telescope ({\it EGRET}) onboard Compton Gamma Ray
Observatory ({\it CGRO}) confirmed 7 gamma-ray pulsars \citep{Thompson04}.
The light curves of these pulsars are characterized by widely separated
double-peak features, with the first peak typically lagging the radio pulse
by a small fraction of rotation period. Recently, {\it AGILE} mission
has detected several new gamma-ray pulsars and refined the detection of
some previously known pulsars \citep{Halpern_etal08,AGILE09a,AGILE09b}.
More impressively, one year after launch, the Large Area Telescope ({\it LAT})
onboard {Fermi Gamma-ray Space Telescope} ({\it Fermi}) has discovered
more than 40 new gamma-ray pulsars \citep{FermiCTA1,Fermi09a,Fermi09b,
FermiRelease09a,FermiRelease09b,FermiPulsars09}. The sensitivity and timing
capability of {\it Fermi-LAT} have also provided the most precise light
curves and phase resolved pulsar spectra \citep{FermiVela,
Fermi09c,Fermi09d}. These new discoveries have greatly expanded the sample
of gamma-ray pulsars and underscored the importance of understanding the
origin of double-peak profiles. At the same time, the diversity of light
curves from latest gamma-ray pulsar sample calls for improvements in
theoretical modeling.

The widely separated double peak feature in the gamma-ray light curves
suggests emission coming from the outer magnetosphere. Gamma rays are
believed to be produced by energetic particles accelerated in the ``gap"
regions in the magnetosphere. The particles emit via curvature, synchrotron
and inverse Compton (IC) radiation. Various theoretical models differ in the
assumed location of the emission zones (i.e., the gaps). Conventional models
for pulsar gamma-ray emission include the polar cap model
\citep{Harding_etal78,DH82,DH96}, the slot gap model (SG, or two-pole caustic
model, TPC for short; \citealp{AS79,Arons83,MH03,MH04a,DyksRudak03,DHR04}),
the outer gap model(OG, \citealp{CHR86a,CHR86b,RY95,Yadi97,CRZ00}), as well
as the inner annular gap model (IAG, \citealp{Qiao_etal04,Qiao_etal07}).

Most calculations from these models approximate the pulsar magnetosphere by
a rotating magnetic dipole field. In reality, the magnetosphere is known to
be filled with plasma \citep{GJ69}. This plasma is essentially force-free
(FF), satisfying $\rho\mb{E+j\times B}/c=0$, where $\rho$ and ${\mb j}$ are
the charge and current densities. The structure of the FF magnetosphere
differs substantially from the rotating dipole field due to the poloidal
current and the returning current sheet
\citep{CKF99,Gruzinov05,Komissarov06,McKinney06,Timokhin06}. Recently, FF
magnetospheric field structure in 3D has become available from the
time-dependent FF simulations \citep{Spitkovsky06,KC09}. The FF magnetosphere
satisfies ${\mb E\cdot \mb B}=0$ everywhere, hence no particle acceleration or
emission can formally occur. Real pulsars must lie somewhere between the vacuum,
which has ${\mb E\cdot \mb B}\ne0$ everywhere, and FF cases. The fact that the
power of gamma-ray emission is a small, though non-negligible ($\lesssim 20\%$),
fraction of the total spin down power suggests that FF should be a reasonable
approximation to the overall field structure with the exception of relatively
small regions where acceleration takes place.

In the companion paper (\citealp{BS09a}, hereafter BS10), we demonstrated that
the modeling of gamma-ray pulsar light curves using the vacuum dipole field is
subject to large uncertainties. This is because the peaks in the vacuum light
curve are caused by the fortuitous overlap of radiation from different regions of
the magnetosphere (caustics). This overlap is sensitive to the geometry of the
magnetic field due to two effects. First, the shape of the field controls how
the aberration of light and light travel delay add together to cause the formation of
caustics. We showed that a seemingly small change of treating the retarded
vacuum field in the instantaneously corotating frame, rather than in the
laboratory frame where it should be defined, can lead to significant reduction
in the strength of caustics in the two-pole caustic/slot gap model and to
the moderate dilution of the outer gap peaks. Second, the behavior of the field near
the light cylinder determines the shape of the polar cap on the star, and, hence,
indirectly controls the shape of the emission zone, even if it is located in the
inner magnetosphere. We considered the sensitivity of the caustic formation to
the modification of shape of the vacuum polar cap. We compared the polar cap
obtained by tracing the vacuum field lines to a simple circular polar cap. This
change significantly affected the caustics of the two pole caustic model. The
outer gap model was less sensitive to this change, but, since the emission zone
for the outer gap is in the outer magnetosphere, the field shape there is likely
to be strongly modified by the inclusion of plasma effects. Given these
uncertainties in the vacuum field modeling, it is highly desirable to revisit
the existing models using the FF field configuration and compare the results
with the latest observations.

In this paper, we use the force-free field from 3D time-dependent
simulations by \citet{Spitkovsky06} (hereafter, S06), and present the
calculation of pulsar gamma-ray light curves using various
theoretical models of emission. In \S2, we provide the general analysis of FF
field structure, addressing the location of the current sheet and its
association with magnetic field lines. In \S3, we describe our numerical
method and theoretical models for calculating the gamma ray light curves.
\S4-\S6 are dedicated to comparison between different models. We show in \S4
that the conventional two-pole caustic model fails to produce double-peak light
curves using the FF field. We propose a novel ``separatrix layer" (SL) model in
\S5, in which the emission zone is fixed in a layer just inside the separatrix
associated with the strong current sheet. This layer can be accurately traced
by open field lines originating in an annulus just inside the polar cap rim\footnote{In
the original version of this manuscript, we termed this model as ``annular gap"
model. The name ``separatrix layer" more accurately describes the location of the
emission zone and avoids the confusion with the ``inner annular gap" model
\citep{Qiao_etal04}. Moreover, although the present paper is based only on
geometry, the association of the emission layer with the magnetospheric current
sheet (see \S\ref{ssec:origin}) implies that the acceleration may not necessarily
involve a ``gap" at all. }. This model differs from the conventional TPC model in that
the emission zone is not concentrated on the last open field lines, but rather interior
to them, and it spans a larger range of heights up to and beyond the light cylinder.
It differs from the conventional OG model in that the emission zone consists of all
field lines in a given flux tube, and not just those that cross the null charge surface.
The double-peak light curve is a generic property of the FF field structure,
where caustics of emission form in the outer magnetosphere when the field
lines approach the split-monopolar geometry. We call this ``the stagnation
effect" on the sky map. In \S6, we show that the conventional OG model applied
to FF field has difficulty in producing double-peak light curves unless a
carefully chosen inclination angle and viewing geometry is used. 
We discuss various implications of our results to the pulsar
acceleration mechanisms and compare our results with recent observations in \S7.
The main results are summarized in \S8.

\section{Structure of the Force-free Magnetosphere}

Force-free pulsar simulations evolve a rotating magnetized conducting sphere
immersed in a massless infinitely conducting fluid. The simulations are
three-dimensional (3D) and start with an inclined dipolar field attached to a 
sphere (S06). The resolution of our simulations is $80$ cells per light cylinder
(LC) radius, $R_{LC}=c/\Omega$, and the radius of the star in simulations is 15
cells ($R_N^{\rm sim}\sim0.19R_{LC}$). We ran a series of simulations varying the
inclination angle between pulsar's magnetic and rotation axes from $\alpha=0^\circ$
to $\alpha=90^\circ$ with an interval of $15^\circ$. The FF fields are extracted
after evolving the FF equations for 1.2 stellar rotations. Evolving for longer times
makes little difference to the field structure \citep{KC09}.
In \S\ref{ssec:rhoj}, we illustrate the charge density and current structure of the FF field,
and present the properties of FF polar cap in \S\ref{ssec:polarcap}. The 3D current sheet
structure of the FF field is discussed in \S\ref{ssec:cusheet}.

\subsection{Charge and Current in the Force-Free Field}\label{ssec:rhoj}

While it is not possible to determine individually the density
and velocity of positively and negatively charged particles
solely from the FF model, the FF current and charge density contain
rich physical information from which we can gain useful insights.

In ideal force-free electrodynamics, by assuming that
electromagnetic field pattern corotates with the pulsar, we
obtain \citep{Gruzinov06}
\begin{equation}
{\mb E}=-{\vec\beta}_0\times{\mb B}\ ,\label{eq:ff1}
\end{equation}
\begin{equation}
\nabla\times[{\mb B}+{\vec\beta}_0\times({\vec\beta}_0\times {\mb
B})]=\lambda{\mb{B}}\ ,\label{eq:ff2}
\end{equation}
where ${\vec\beta}_0\equiv{\mb\Omega}\times{\mb r}/c$ is the
normalized corotation velocity, and $\lambda$ is a scalar function that
is conserved along the magnetic field lines, because
$\mb{B}\cdot\nabla\lambda=0$. In the case of the aligned rotator,
parameter $\lambda$ is the analog of $A'(\Psi)$ in the pulsar
equation \citep{Michel73a,Michel73b}, where $A$ is proportional to
the poloidal current, and the derivative is with respect to the poloidal
magnetic flux $\Psi$. Therefore, $\lambda\propto A'$ characterizes
the current per magnetic flux. More specifically, positive $\lambda$
represents current along the magnetic field line, while negative
$\lambda$ represents current flow opposite to the direction of the
magnetic field. A formal derivation for the physical meaning of $\lambda$
can be found in Appendix \ref{app:derivation}, where we show that $\lambda {\bf B}$ is 
the field-aligned current density in the corotating frame. 
We calculate
$\lambda$ by taking the dot product of ${\mb B}/B^2$ on both sides of
equation (\ref{eq:ff2}).

\begin{figure*}
    \centering
    \includegraphics[width=160mm,height=60mm]{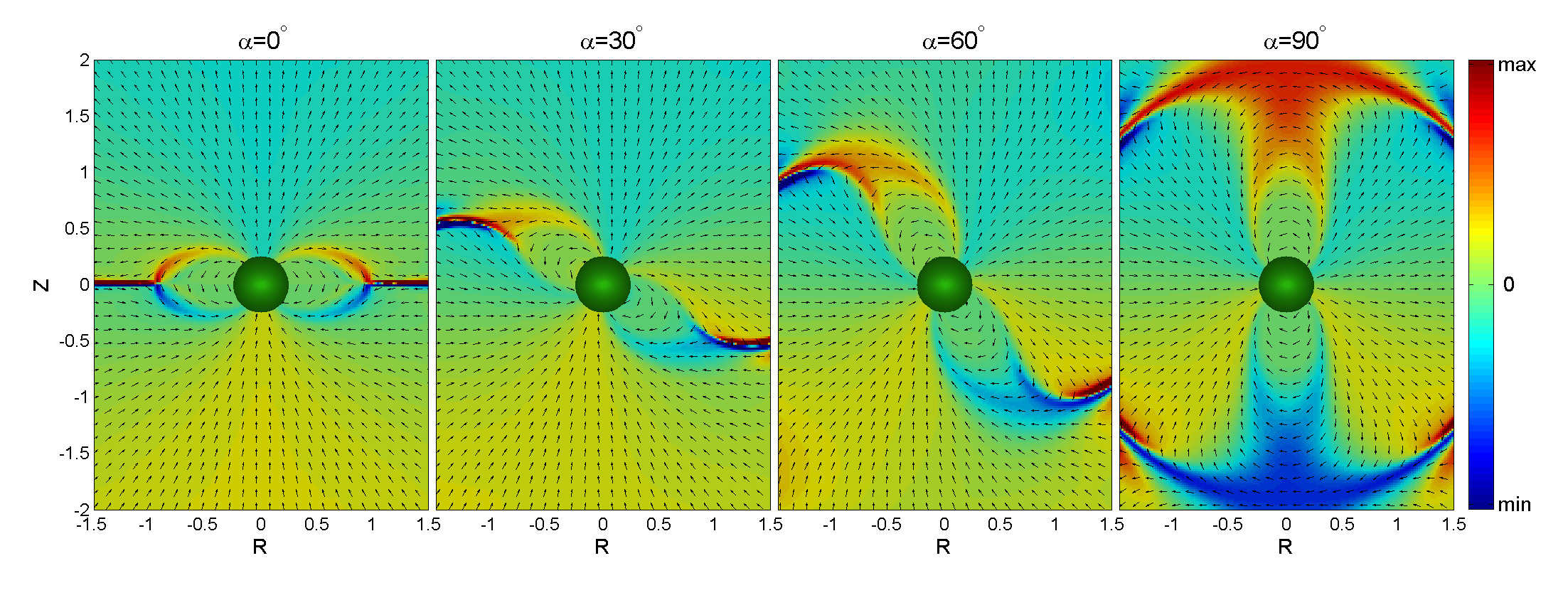}
  \caption{Color plots of the parameter $\lambda$ of the force-free field in the
  $\Omega-\mu$ plane for magnetic inclination angles $\alpha=0^\circ, 30^\circ, 60^\circ$
  and $90^\circ$ respectively. Arrows show the direction of the projected
  magnetic field in this plane. All plots have units in light cylinder radius
  $R_{LC}$, and have the same color scale.}\label{fig:lamB}
\end{figure*}

In Figure \ref{fig:lamB}, we show the color plots of $\lambda$ in the $\Omega-\mu$
plane (the plane containing the rotation axis and the magnetic moment vector) for
a range of inclination angles. We also show vectors of the magnetic field projected
onto this plane. The strong current layer manifests itself as a local enhancement of
$\lambda$. First, consider the axisymmetric pulsar magnetosphere ($\alpha=0$).
The field structure closely resembles the solution by \citet{Timokhin06} with
$x_0\equiv R_Y/R_{LC}\lesssim1$ and well approaches the CKF solution\footnote{In
our simulation, the numerical resistivity is very low when the equatorial current sheet is
aligned with the grid. Therefore, for inclination angles $<10^\circ$ the evolution of the
Y-point to the light cylinder takes longer than $2$ turns to approach the expected result
(S06). Larger inclination angles are well converged by $1.2$ turns. We have checked
this convergence by comparing results between 1.2 turns and 2 turns (where we manage
to avoid reflection by using a very large computational domain). No significant difference
is found for $\alpha>15^\circ$. In Figures 1-4, however, we do show the field structure
after 2 turns, to allow the Y-point in the $\alpha=0^\circ$ case to get closer to the LC.} \citep{CKF99}.
In agreement with these solutions, the current along the open field lines in our
simulation is predominantly of one sign in each hemisphere; however, near the edge
of the open zone there is a distributed return current region bounded by a
strong return current sheet. Instead of a delta-function current sheet expected
in theory, the simulation current sheet is several grid cells thick, giving
$\lambda$ a finite maximum/minimum there. The open field lines increasingly
resemble the rotating split monopole solution beyond the LC and become radial
in the poloidal section. The left panel of Figure \ref{fig:lamB} illustrates
these features.

Next, consider the general oblique rotators. The transition of current structure is
smooth as $\alpha$ increases from $0^\circ$ to $90^\circ$. In Figure \ref{fig:lamB},
we see that at small inclination angle, $\alpha=30^\circ$, the pattern of $\lambda$
is similar to the axisymmetric case, and the current sheet inside the LC appears
slightly wider and weaker. As the inclination angle increases, there are also current
sheet-like features inside the LC, but the distribution of current becomes very 
asymmetric, and more current is returned to the polar cap through the distributed
flow rather than the current sheet. The current flow in the polar cap is predominantly
of one sign (e.g., ingoing) for the aligned rotator, bounded by a strong current sheet
of opposite sign (e.g., outgoing), and is symmetric with respect to the equatorial
plane. In contrast, when $\alpha=90^\circ$, the current flow has different signs
in the northern and southern halves of the polar cap (this can be better seen in
Figure \ref{fig:polarcap}, which shows the current through the polar cap, see
\S\ref{ssec:polarcap}), 
and the current distribution is anti-symmetric with respect to
the equatorial plane. On the periphery of the polar cap there are thin current
layers, as shown in red (upper) and blue (lower) oval  structures in the right panel
of Figure \ref{fig:lamB}. The current in these current layers forms loops connecting
the two poles through the closed zone. Integrating over the polar cap, we find that
the amount of current flowing to the other pole is about $20\%$ of the current flowing
on open field lines. Such closed current loops for orthogonal rotators are qualitatively
consistent with predictions of the model by \citet{Gurevich_etal93}, although we find
the amount of current shunting to the other pole to be smaller in our simulations. The
total current on the open field lines (integrated by magnitude) is just $20\%$ smaller 
for the orthogonal rotator than for the aligned rotator, suggesting that the current
density for the orthogonal rotator exceeds the simple expectation of the speed of
light times local Goldreich-Julian density on the polar cap. This current cannot be
provided by charge-separated flow alone and requires abundant pair formation.
The details of the current adjustment in such polar caps are still uncertain
\citep{Lyubarsky09}.

A thin current sheet {\it outside} the LC exists for all inclination angles; its structure
asymptotically approaches the rotating split monopole solution outlined by
\citet{Bogovalov99}. If the current sheet outside the LC is connected to the star, it must
be connected through a current sheet inside the LC, because the current flow in the CF
can not cross magnetic field lines, as inferred from equation (\ref{eq:lambda2}). In Appendix
\ref{app:cusheet}, we show that the amount of current in the current sheet outside the LC
that is connected to the star monotonically decreases with inclination angle. This is related
to the degradation of the current sheet {\it inside} the LC and the thickening of the Y-region
for oblique rotators shown in Figure \ref{fig:lamB}. These features are unlikely
to be caused by numerical resistivity, and we have tested that the thickness of the Y-region
and the strong current layers inside the LC is not sensitive to numerical resolution. For the
orthogonal rotator, the current sheet outside the LC is totally disconnected from the star
(see Appendix \ref{app:cusheet}). Therefore, the oval structures in the right panel of Figure
\ref{fig:lamB} are most likely the current loops that connect the two poles, rather than
current sheets connecting to the outer magnetosphere. This strongly contrasts with the case
of the aligned rotator, where the current sheet inside the LC unambiguously exists and
connects to the equatorial current sheet outside the LC. While there are definitely strong
currents flowing on the periphery of the open field lines for all other inclinations seen from
Figure \ref{fig:lamB}, determining whether their thickness is finite or infinitesimal would
require further study (see \S\ref{ssec:polarcap} for a speculative discussion). Throughout
this paper, we will refer to these current sheet-like features inside the LC as ``strong current
layers".

\begin{figure*}
    \centering
    \includegraphics[width=160mm,height=60mm]{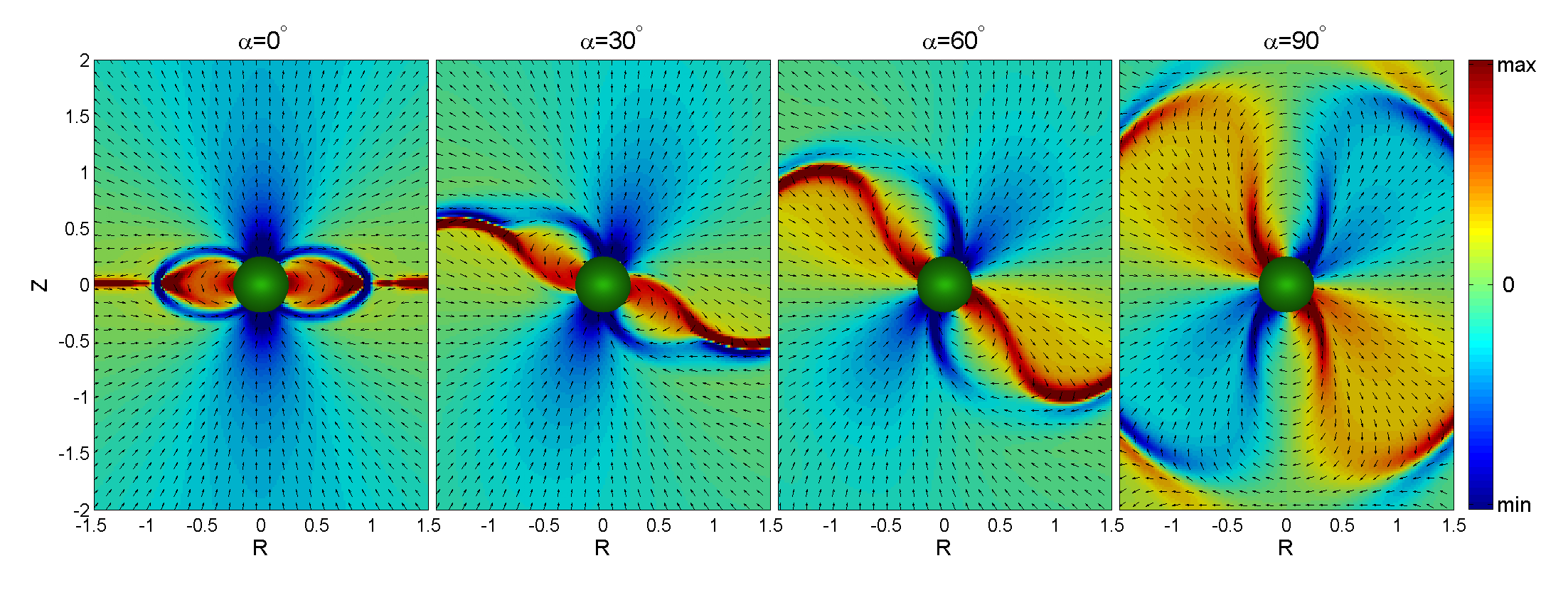}
  \caption{Color plots of the charge density $\rho$ in the force-free field in
  the $\Omega-\mu$ plane for magnetic inclination angles $\alpha=0^\circ, 30^\circ, 60^\circ$
  and $90^\circ$. We've multiplied $\rho$ by $r^2$ to improve the contrast. Arrows
  indicate the direction of the projected magnetic field in this plane. All plots
  have units in light cylinder radius $R_{LC}$, and have the same color scale.}\label{fig:rhoB}
\end{figure*}

\begin{figure*}
    \centering
    \includegraphics[width=160mm,height=60mm]{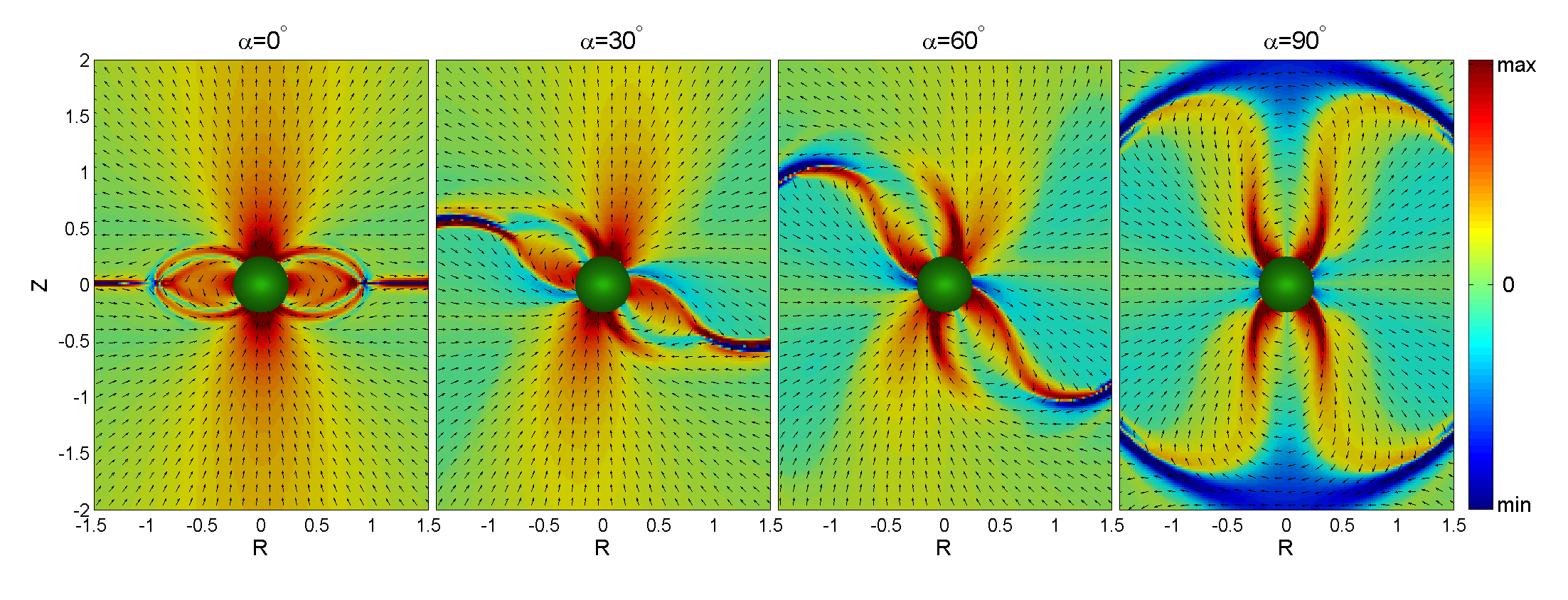}
  \caption{Color plots of the parameter $\varrho$ [see eq. (\ref{eq:netj})] of the
  force-free field in the $\Omega-B$ plane for magnetic inclination angle
  $\alpha=0^\circ, 30^\circ, 60^\circ$ and $90^\circ$. We've multiplied $\varrho$
  by $r^2$ to improve the contrast. Arrows show the direction of the projected
  magnetic field in this plane. All plots have units in light cylinder radius
  $R_{LC}$, and have the same color scale.}\label{fig:netjB}
\end{figure*}

The charge density necessary to provide corotation of the magnetosphere is referred
to as Goldreich-Julian (GJ) charge density \citep{GJ69}. It can be written as
\begin{equation}
\rho=\frac{\nabla\cdot{\mb E}}{4\pi}=-\frac{{\mb\Omega}\cdot{\mb
B}}{2\pi c}+\frac{{\vec\beta}_0\cdot(\nabla\times{\mb B})}{4\pi}\
.\label{eq:rho}
\end{equation}
If we assume complete charge separation, as in \citet{GJ69}, then the second term
is of the order $\rho\beta_0^2$, and we arrive at the classical result with
$\rho\propto{\mb\Omega}\cdot {\mb B}$. Whether a real pulsar magnetosphere has
complete charge separation is not clear, but the eclipse in the double pulsar
system PSR J0737-3039 indicates that the particle density in the inner
magnetosphere is much higher than typical Goldreich-Julian density
\citep{RafikovGoldreich05,LyutikovThompson05ApJ}. In Figure \ref{fig:rhoB}, we
plot the charge density from the FF simulation. We find that the bulk
distribution of FF charge density is approximately proportional to
${\mb\Omega}\cdot {\mb B}$. However, there is a significant enhancement of
charge density in the strong current layer and in the current sheet beyond the
LC, which is due to the contribution of the current in the second term of equation
(\ref{eq:rho}).

The distribution of charge in the strong current layer inside the LC
has another notable feature. In Figure \ref{fig:rhoB}, the charge
distribution is symmetric between the north and south magnetic hemispheres
for the aligned rotator. Dense negative charges more or less uniformly
fill the current sheet inside the Y-point, while the current sheet
outside the Y-point is positively charged \citep{Timokhin06}. However, this
symmetry is broken when $\alpha$ is different from zero (e.g., at
$\alpha=30^\circ$). The strong current layer inside the LC then contains both
signs of charge, with the northern and southern parts having opposite sign at
fixed azimuth. The current sheet outside the LC also contains charges of both
signs, although one sign of charge is prevailing (in red color). This trend
is also observed by \citet{CK10}.

It is also useful to visualize the FF magnetosphere in terms of
space-like and time-like regions of 4-current.  
A 4-current is space-like if ${\mb j}^2-\rho^2>0$; otherwise, it is
time-like. Note that this definition is Lorentz invariant. 
Regions with
space-like current demand counter-streaming of different signs of charge in the
frame where $\rho=0$. This could lead to plasma instabilities and dissipation
\citep{Lyubarsky96, Gruzinov07b}. A current sheet with space-like current could then be
expected to produce particle acceleration and gamma-rays
\citep{Gruzinov07a,Lyubarsky08,Spitkovsky08}. In the asymptotic split-monopole regime of
\citet{Bogovalov99}, one expects the current sheet outside the LC to be
space-like. This is because both magnetic and electric fields change sign
across the current sheet, and since $|{\mb B}|>|{\mb E}|$ outside the current
sheet, MHD jump condition demands $J>\sigma c$ in the current sheet, where $J$
and $\sigma$ are surface current and surface charge densities\footnote{We thank Yuri
Lyubarsky for pointing this out to us.}. In Figure
\ref{fig:netjB}, we plot the Lorentz invariant quantity
\begin{equation}
\varrho\equiv\pm\sqrt{|\rho^2c^2-{\mb j}^2|}\ ,\label{eq:netj}
\end{equation}
where we take plus sign for time-like current, and minus sign for
space-like current; thus, counter-streaming is required in the
regions which are negative. We find that in both aligned and oblique
rotators the current sheet outside the LC is in general space-like (blue).
Although there are regions in red around the current sheet, the overall
surface integrated $\varrho$ is negative, as expected. The strong current
layer inside the LC appears to be predominantly time-like. The magnetosphere
in the aligned rotator is dominated by time-like regions, and only in regions
near the null charge surface does it show very small negative values of
$\varrho$. As the inclination angle increases, the space-like region becomes
much broader, occupying a substantial fraction of the open field-line regions.
We will discuss this further in \S\ref{sec:discussion}.

\subsection[]{The Force-free Polar Cap}\label{ssec:polarcap}

\begin{figure*}
    \centering
    \includegraphics[width=160mm,height=54mm]{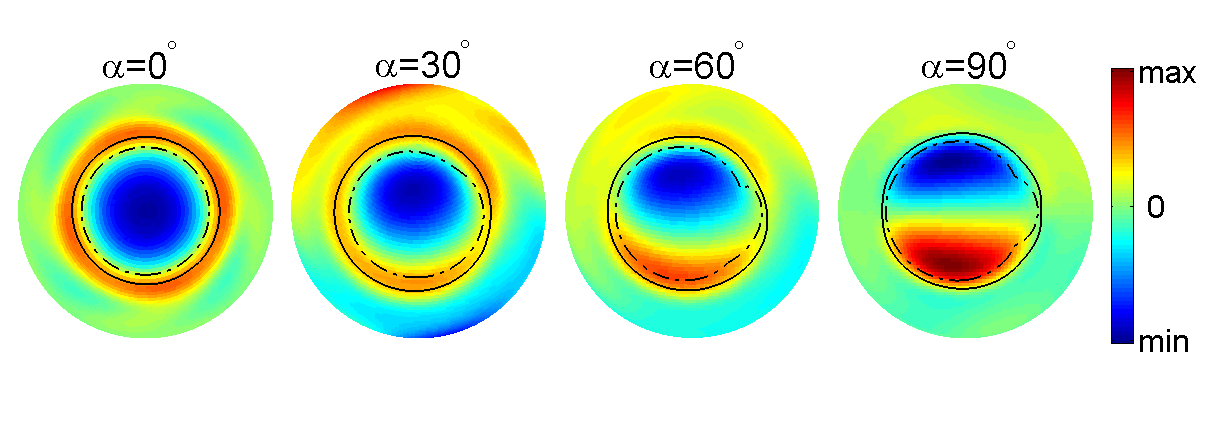}
  \caption{The shape of the polar caps in force-free field (solid lines)
  and the Deutsch field (dash-dotted lines) for magnetic inclination
  angles $\alpha=0^\circ, 30^\circ, 60^\circ$ and $90^\circ$ (left to right).
  Overlayed are the color plots of the current to flux ratio $\lambda$
  on the pulsar surface. To better visualize $\lambda$,
  the size of the pulsar is set to be $0.25R_{LC}$. Also, the color scale
  of $\lambda$ is compressed compared to Figure \ref{fig:lamB}.}\label{fig:polarcap}
\end{figure*}

The polar cap is the region on the neutron star (NS) surface where open
magnetic field lines originate. The rim of the polar cap is set by
the locus of last open field lines (LOFLs). We adopt the conventional
definition of LOFLs, where we treat a field line as open if it crosses
the LC; otherwise, it is regarded as closed. This is an oversimplification,
as in FF field the closed zone does not have to extend to the LC everywhere;
some closed field lines may also extend through the LC and close through the
current sheet. However, this simple definition facilitates comparison with
calculations using the vacuum field\footnote{Moreover, it is probably better
not to distinguish between open and closed field lines, but to distinguish
between ``active" and ``dead" field lines. Dead
field lines are closed within the Y-point, beyond which the equatorial current
sheet is launched. Other field lines, including open field lines and field
lines that are closed through the current sheet, are considered active.
Therefore, one may consider ``last active field
lines" as a more meaningful concept in the FF magnetosphere. Our
definition of last open field lines as active field lines is equivalent to setting the
radius of the Y-point to be at the light cylinder.}. In Figure
\ref{fig:polarcap}, we plot the shape of the FF polar cap on the sphere of
radius $0.25R_{LC}$ for inclination angles $\alpha=0^\circ, 30^\circ,
60^\circ, 90^\circ$. The surface of the sphere (which is larger than the NS
for ease of visualization) is colored with the parameter $\lambda$,
indicating the sense of the current flow. For comparison, we also plot with
dash-dotted lines the polar caps obtained from the retarded vacuum magnetic
dipole field \citep{Deutsch55}. We note that tracing vacuum field lines
beyond the LC always returns them back to the star, because vacuum solution
does not have the current sheet. Nevertheless, as is conventional in the
literature, we always call vacuum field lines that go beyond the LC ``open",
although they are still formally closed.

From Figure \ref{fig:polarcap} we see that the polar cap of the FF field
encloses the current flowing into and out of the pulsar. The polar current
switches from flowing into the NS at $\alpha=0^\circ$, to equal halves of
oppositely-directed current at $\alpha=90^\circ$. The peak of the current
density slightly lags the zero phase (i.e., the phase of the magnetic pole)
due to pulsar rotation. The evolution of the strong current layer as
$\alpha$ increases can also be clearly seen from Figure \ref{fig:polarcap}.
For aligned rotator, the rim of the polar cap corresponds to the footprint of
the current sheet (i.e., the red enhancement on the left panel), as expected.
As $\alpha$ increases, the strong current thickens on one side (lower side
in Figure \ref{fig:polarcap}), and weakens on the other side. The thickened
part may no longer count as a current sheet, as it gradually occupies half
of the polar cap to become the main contributor of the polar current at
$\alpha=90^\circ$. The weakened side gradually shifts to the outside of the
polar cap (at $\alpha=90^\circ$), and forms the current loop in the closed
field lines as seen in the rightmost panel of Figure \ref{fig:lamB}. Following
the discussion in the previous subsection, we speculate that the weakened
part of the strong current layer actually corresponds to the current sheet
inside the LC. The current sheet covers a circle on the polar cap for the aligned
rotator. For oblique rotators, the region covered by the current sheet on the
polar cap reduces to an arc covering the upper part of the polar cap in Figure
\ref{fig:polarcap}. The extent of the arc-like region shrinks with inclination
angle $\alpha$, consistent with the reduction of star-connecting current in the
current sheet {\it outside} the LC. In the mean time, some current in the arc-like
region on the polar cap becomes connected to the other pole, forming current
loops. This is associated with the thickening of the
Y-region discussed in the previous subsection. The loop current is zero for
the aligned rotator, and increases to about $20\%$ of the total polar current
for the orthogonal rotator.

We note that the FF polar caps are smooth and approximately circular. On
the other hand, polar caps obtained from oblique vacuum dipole rotators
display notch structures and sharp jumps \citep{DHR04}. As was shown in
BS10, gamma-ray light curves are sensitive to the shape of the polar caps,
as this changes the formation of caustics in the emission. The FF polar
cap is also larger than its vacuum counterpart, due to the larger open
magnetic flux. This is consistent with the larger spin down power for
force-free rotator compared to equivalent vacuum rotator (S06). The same
effect is seen in the aligned rotator \citep{CKF99}. As a result, the LOFLs
predicted from the vacuum dipole field should lie within the LOFLs, or in
the open volume, of the FF field. This effect also has significant
consequences to the appearance of the FF sky map.

The magnetic field lines that originate near the polar cap from NS
surface can be parameterized using the open volume coordinate system
\citep{Yadi97,CRZ00,DHR04}. This system has two coordinates, ($r_{\rm ov}$,
$\phi_m$), recording the footprint on the NS surface, where $\phi_m$ is the
magnetic azimuth, and $r_{\rm ov}$ is the magnetic colatitude normalized by
the magnetic colatitude of the LOFLs. Therefore, the LOFLs correspond to
$r_{\rm ov}=1$, and open field lines have $r_{\rm ov}<1$.

\subsection{Current Sheet-Field Line Association}\label{ssec:cusheet}

\begin{figure}
    \centering
    \subfigure{
    \includegraphics[width=80mm,height=80mm]{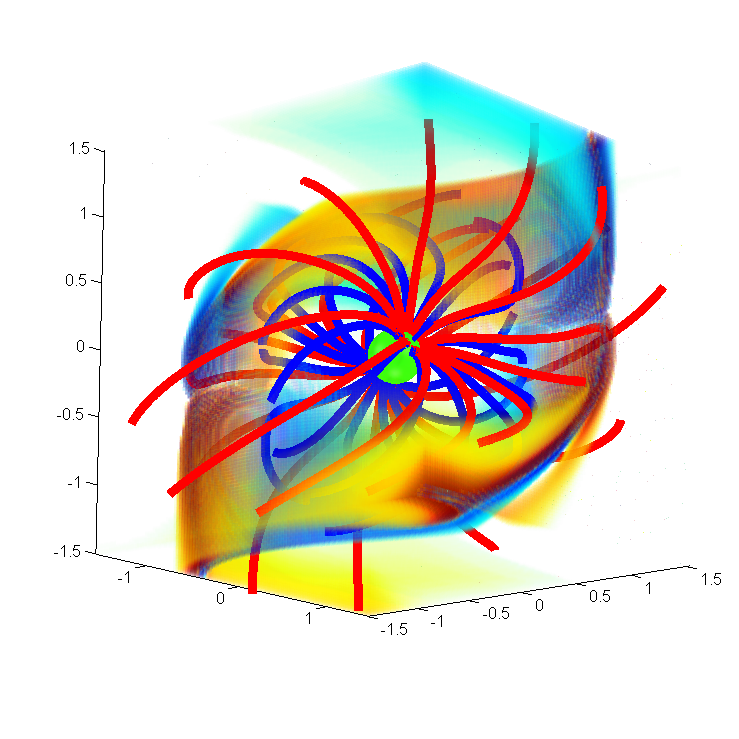}}
    \subfigure{
    \includegraphics[width=80mm,height=80mm]{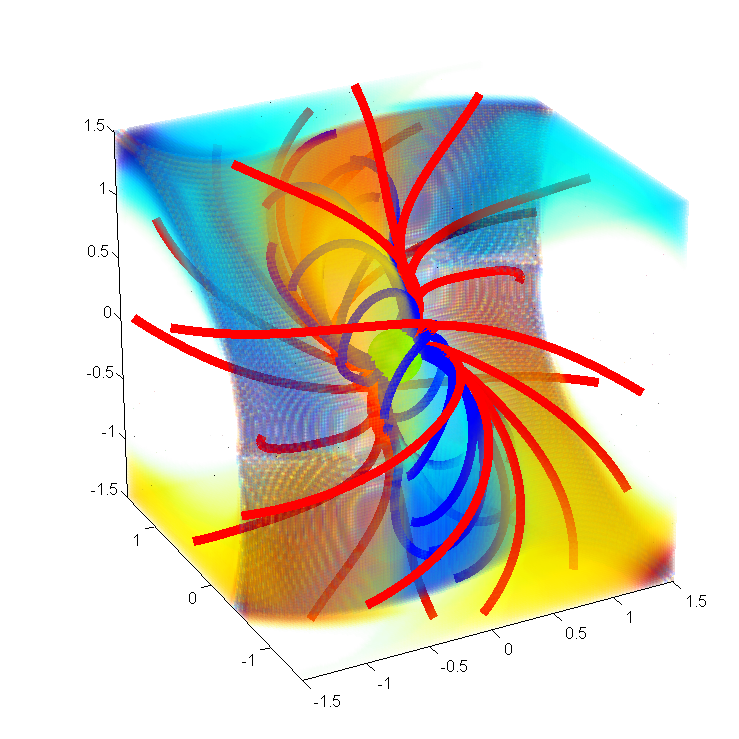}}
  \caption{3D volume rendering of the current to flux ratio $\lambda$, defined
  in equation (\ref{eq:ff2}), for the FF simulation with $\alpha=60^\circ$.
  Transparency is  determined by $|\lambda|$, where only regions with large
  $|\lambda|$ are opaque. The color scale is the same as in Figure \ref{fig:lamB}.
  Blue curves indicate the LOFLs, while the red curves are field lines with
  $r_{\rm ov}=0.90$. The two panels show different viewing angles. Axes are in units
  of $R_{LC}$. The movie version of this plot is available in the online version of
  the paper and at
  {\texttt{http://www.astro.princeton.edu/$\sim$anatoly/cursheet.mpeg}} }\label{fig:cusheet}
\end{figure}

Current sheets are generic features of FF pulsar magnetosphere.
To better visualize the current sheet structure, we present a volume
rendering plot in Figure \ref{fig:cusheet}, showing the magnitude of the
current function $\lambda$ as volume opacity and two sets of flux tubes
($r_{\rm ov}=1$ and $0.9$) as color lines. This figure demonstrates the
relation between regions of strong current and magnetic field lines for
$\alpha=60^\circ$ rotator.

First of all, we see that LOFLs are associated with the strong current layer inside
the LC. Note that the current layer has finite thickness, and it encloses the LOFLs.
Some field lines with $r_{\rm ov}<1.0$ and $r_{\rm ov}>1.0$ also lie within the
strong current layer. Moreover, due to magnetic reconnection in the current sheet (outside
the LC), field lines can be closed beyond the LC, and we observe that some outgoing
field lines with $r_{\rm ov}\gtrsim0.95$ stall and turn inward beyond the LC. These
field lines trace the current sheet outside the LC to a finite distance before
turning back. This effect makes finding the true LOFLs a more complicated exercise
than we perform in this paper.

Secondly, we see that field lines with $r_{\rm ov}\lesssim0.95$ are generally
open in the physical sense. Red curves in Figure \ref{fig:cusheet} show some
representative field lines with $r_{\rm ov}=0.90$. These field lines lie at
the edge of the strong current layer inside the LC, and follow closely
the current sheet outside the LC. The current sheet outside
the LC separates field lines originating from different poles.

We conclude that $r_{\rm ov}$ is a good tracer of current in
the FF magnetosphere. For $\alpha=60^\circ$ rotator, strong current is
associated with $0.95\lesssim r_{\rm ov}\lesssim1.05$, and field lines with
$r_{\rm ov}\lesssim0.95$ lie outside the current sheet. For other
inclination angles, the critical value of $r_{\rm ov}=0.95$ varies, but
not significantly.

\section[]{Emission Calculation}\label{sec:model}

The FF magnetosphere has no intrinsic acceleration and emission since the
condition ${\mb E}\cdot{\mb B}=0$ is satisfied everywhere by construction.
Therefore, we will calculate the pulsar gamma-ray emission based solely on
geometric grounds. Specifically, the calculation involves a prescription of
the emission zone and a method to determine the intensity and direction of
emission inside the emission zone.

\subsection[]{Emission Zone Geometry}

Here we summarize the emission zone geometry prescriptions by different
theoretical models that we consider in this paper.

The two-pole caustic (TPC) model is an extended version of the slot-gap
model (e.g., \citealp{Arons83}). In this model, the emission is assumed to
originate on LOFLs (i.e., $r_{\rm ov}=1$) from the NS surface up to some
cutoff radius \citep{DyksRudak03,DHR04}. We use two cutoff radii: spherical
radius $r_{\rm max}$, measured from the center of NS, and cylindrical radius
$R_{\rm max}$, measured from the rotation axis. Together, these two parameters
define the extent of the TPC emission zone.

In this paper, we also propose a new emission model which we term the
``separatrix layer" (SL) model. In this model, we assume that emission comes
from a layer in the vicinity of the separatrix, which separates open and closed
field lines as well as field lines with different polaralities. As we have
seen in \S\ref{ssec:cusheet}, the location of the separatrix corresponds to
strong current layer/current sheet. This separatrix layer is described by
$r_{\rm ov}\simeq0.90-0.95$, going from the stellar surface up to a cutoff
radius in a way that is similar to the TPC model. The cutoff radii for SL model are not 
limited to be inside the light cylinder. We will discuss the physical
justification for this model in \S\ref{sec:discussion}.

For the outer gap (OG) model, the emission is assumed to originate on
the open field lines beyond the null charge surface (NCS), and extends up
to the LC. Recent refinements of the OG models allow the inner boundary
of the gamma-ray emission to be inside the NCS
\citep{Hirotani07,TakataChang07,TCC07,TCS08}. In order to incorporate
such possibility, we assume that the emission comes from all field lines in
a given flux tube that cross the NCS, extending from the stellar surface to
certain cut-off radius. This approach maximizes the volume of the emission
zone. The emission from the OG model is often assumed to be centered at
certain $r_{\rm ov}$ (e.g., $r_{\rm ov}=0.90$).

Unlike the OG model, the inner annular gap (IAG) model assumes that
particles are accelerated in an annular open field line region extending
from the NS surface to the NCS \citep{Qiao_etal04,Qiao_etal07}. The emission
zone is bounded by LOFLs and the critical field lines, which
cross the LC at the NCS. To simplify, we may also think of emission
from the IAG to be centered at certain $r_{\rm ov}<1$.

For all models mentioned above, we consider the emission originating from
annular regions centered at fixed $r_{\rm ov}$ with a Gaussian width
$\Delta r_{\rm ov}=0.025$ similar to \citet{DHR04} as our standard emission
zone prescription. The radial extent of the emission zone is controlled
by the location of the NCS (for OG and IAG models), as well as the
$R_{\rm max}$ parameter (we do not use $r_{\rm max}$ in this paper).

\subsection[]{Calculation Method}

In calculating the gamma-ray light curves we adopt two basic
assumptions. First, we assume that the emission is produced by outgoing
energetic particles in the emission zone. These particles are confined
to travel almost along the magnetic field line, but may acquire some
pitch angle $\theta_p$. Such pitch angle is expected from both conventional
SG and OG models, e.g., acquired during the pair creation process \citep{Tang_etal08},
or via the cyclotron resonant absorption \citep{Harding_etal08}. If emission
originates from the current sheet, as we postulate in \S\ref{ssec:origin},
we would expect even larger pitch angles since in the current sheet, magnetic
field is relatively weak while the plasma is likely to be hot due to reconnection.
The emission mechanism can involve synchrotron, curvature, and inverse-Compton,
but the direction of the emitted photon is along the direction of particle motion
due to relativistic beaming. Calculation of the direction of emitted photon in
the lab frame (LF), including the aberration and time delay effects, is discussed
in detail in BS10. Here, we generalize these results to incorporate the effect
of  finite particle pitch angle. Consider a drift frame (DF) in which the
electric field vanishes. Let the velocity of the DF relative to the LF be
${\mb V}_{d}$ and $\vec{\beta}_{d}={\mb V}_{d}/c$. In this frame, let $\theta_p$
be the particle pitch angle. Then the emission direction is determined by
\begin{equation}
{\mb e}=(\alpha\cos\theta_p){\mb t}+(\alpha\sin\theta_p\cos\varphi){\mb n}
+(\alpha\sin\theta_p\sin\varphi){\mb b}+\vec\beta_d\ ,\label{eq:aberration}
\end{equation}
where ${\mb t}$, ${\mb n}$, ${\mb b}$ are the tangential, normal and binormal
unit vectors to the magnetic field line (in the LF), and $\alpha$ is determined
by requiring that $|{\mb e}|=1$.

This formula is essentially the same as equation (6) of \citet{TCC07},
except that it applies in any drift frame. The drift velocity can be chosen
to be either the corotation velocity $\vec\beta_0$, or the ${\mb E}\times{\mb B}$
drift velocity. When $\theta_p=0$, equation (\ref{eq:ff1}) ensures that these
two choices produce the same result. However, this is no longer the case for
finite pitch angle. This is because pitch angle is not Lorentz invariant and
depends on frames. Moreover, the drift frame velocity with $\vec\beta_d=\vec\beta_0$
is no longer valid at $R>R_{LC}$ since the frame velocity becomes superluminal.
Therefore, a better choice is to use the ${\mb E}\times{\mb B}$ drift velocity,
which is the minimum drift velocity among all DFs, and this is the drift velocity
we adopt throughout this paper. In our calculation, we further assume that the
pitch angle is constant for all particles in the emission zone.

Our second assumption is constant emissivity along particle trajectories
in the emission zone. In reality, as long as the particle emissivity does
not vary dramatically in the emission zone, our calculation will be able
to catch the overall features of gamma-ray light curves. Moreover,
as pointed out in BS10, the overall appearance of the emission sky map
is more sensitive to field structure, than to emissivity, because the
formation of caustics largely depends on the details of field configuration.
Adopting constant particle emissivity will help us locate possible emission
zones in FF field by comparing with observations. For example, comparing
model light curves with observations can help identify the radial extent
of the emission zone (i.e., $R_{\rm max}$, see \S\ref{sec:AG}), which then
places constraint on the physics of radiation mechanism.

One basic difference in our approach compared to previous work (e.g.,
\citealp{Yadi97,CRZ00,DHR04,TakataChang07,Harding_etal08,Watters09}) is that
the emission is weighted by the length along particle trajectories rather than
along magnetic field lines.
 We've discussed in BS10 that particle trajectory
follows magnetic field lines in the corotating frame (CF), and the magnetic
field in the CF is the same as lab frame (LF) magnetic field.
Therefore, to calculate particle trajectories, we trace magnetic field lines in
the LF, and add the effect of stellar rotation. In Appendix \ref{app:geometry}
we provide a simple relation between segment length along a magnetic field line
and along the corresponding particle trajectory. The difference is small well
inside the LC, but the correction becomes important near the LC and beyond. Since
FF field does not break down at the LC, we will allow the emission zone to extend
beyond the LC, where such correction is necessary\footnote{For example, consider
a rotating split monopole field, where particle trajectories are straight lines,
while the field lines are spirals. The FF field asymptotically approaches the split
monopole field, and such correction avoids overweighing far-field emission.}.

Finally, in our calculation photons from the emission zone are projected to
the sky map ($\phi, \xi_{\rm obs}$), where $\phi$ corresponds to the rotation
phase, and $\xi_{\rm obs}$ denotes the observer's viewing angle. A time delay
correction is applied during the projection as usual
\begin{equation}
\phi=-\phi_e-{\mb r}\cdot{\mb e}/R_{LC}\ .\label{eq:tdelay}
\end{equation}

The brightness of the sky map is proportional to the number of photons projected
to the sky map bins, divided by the solid angle of each bin. The solid angle of
each sky map bin is $d\Omega=\sin\xi_{\rm obs}d\xi_{\rm obs}d\phi$; therefore,
the sky map brightness is proportional to photon counts in each bin divided by
$\sin\xi_{\rm obs}$. Cutting the sky map at $\xi_{\rm obs}$ produces the light
curve seen by the observer. The correction of $1/\sin\xi_{\rm obs}$ only affects
the normalization of the light curves at each observer viewing angle, but not the
shape of the light curves. Including it is important for the calculation of flux
correction factor used to infer the intrinsic luminosity of gamma-ray pulsars.

\section[]{The Two-Pole Caustic Model}\label{sec:tpc}

\begin{figure*}
    \centering
    \includegraphics[width=170mm,height=170mm]{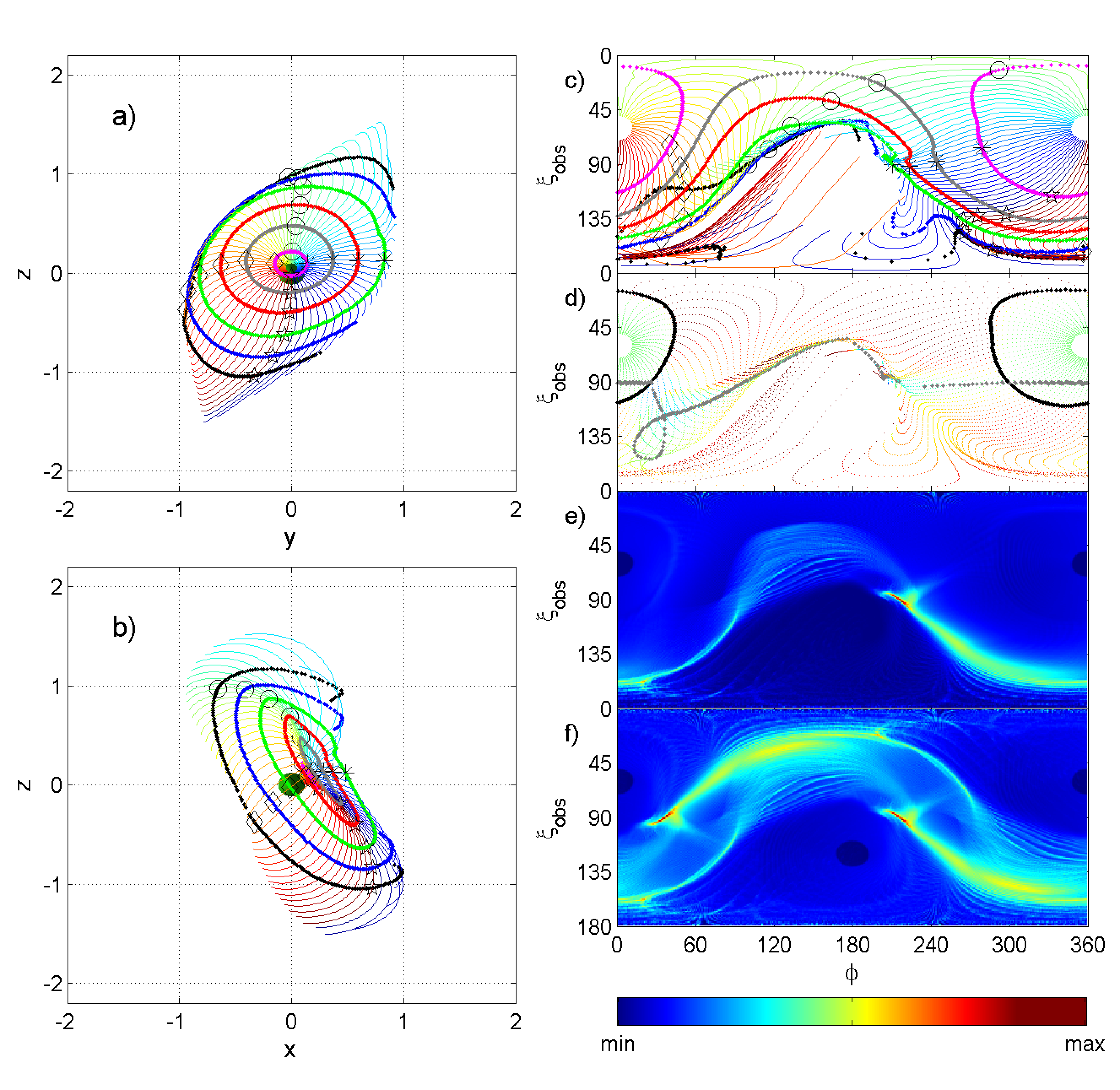}
  \caption{Sky map and field line structure from the TPC model using FF field.
  The inclination angle is $\alpha=60^\circ$. All field lines have $r_{\rm ov}=1.0$
  and are traced to the LC. a) and b): Field lines from the north pole viewed from
  two different directions. c): Projection of field lines from the north pole onto
  the sky map. Line colors are same as in a,b). Color rings represent points located
  at the same distance along the magnetic field lines from the center of NS: pink
  (0.25$R_{LC}$), grey (0.50$R_{LC}$), red (0.75$R_{LC}$), green (1.0$R_{LC}$), blue
  (1.25$R_{LC}$), black (1.50$R_{LC}$). Asterisks, circles, diamonds and stars are
  added to help in identifying field lines. d): Line plot of the sky map with each
  point colored by 4-current length $\varrho$ [see equation (\ref{eq:netj})];
  $\varrho=0$ corresponds to the middle of the color bar. NCS is indicated by the
  grey dots, and the black circles correspond to the actual size of the star in the
  simulation ($R_N^{\rm sim}\sim0.19R_{LC}$). e): Sky map from field lines from the
  north magnetic pole. f) Sky map from both poles. In e) and f), zero brightness
  corresponds to the dark blue as indicated by ``min" in the color bar. See text for
  more details.}\label{fig:TPCSM60_0}
\end{figure*}

In this section we consider the emission from TPC model ($r_{\rm ov}=1.0$) in FF
field. The 3D shape of the last open field lines and the corresponding sky maps are
shown in Figure \ref{fig:TPCSM60_0} for inclination $\alpha=60^\circ$. In order to
better see the origin of the sky map features, we mark different field lines with
different colors in panels (a) and (b), and use the same color convention to plot the
sky map in panel (c). To guide the eye, we also plot six color rings crossing the field
lines in panels (a), (b) and their projections onto the sky map in panel (c). Points on
each ring are located at fixed distance, $s$, from the center of NS, measured {\it along}
the field lines. The six rings span from $s=0.25R_{LC}$ to $1.50R_{LC}$ with an interval
of $0.25R_{LC}$. As the size of the star in the simulation is relatively large,
$R_N^{\rm sim}=0.19 R_{LC}$, tracing emission from the surface out would  leave large
gaps on the sky map surrounding the poles. To fill these gaps, we add tracing of radiation
from inside the star, where we impose a vacuum dipole field down to the radius
$R_N=0.01 R_{LC}$. This makes for effective size of the star for the purposes of sky map
calculation to be $R_N$. As  FF field is nearly dipolar close to the star, the matching to
the vacuum field is easy to achieve. In panel (d), we make the line plot of the sky map as
in panel (c), but the lines are colored with the value of 4-current length $\varrho$,
defined in equation (\ref{eq:netj}). Since current drops rapidly with radius, the actual
color scale is set by $\varrho r^2$. Note that in this panel, $\varrho=0$ corresponds to
the middle of the color bar. The black semi-circles around the pole indicate the size of
the star in the simulation $R_N^{\rm sim}$. At smaller radii, where the FF field is not
available, $\varrho$ is taken to be zero. Panels (e) and (f) display the brightness of the
sky map. Note that panel (e) shows the sky map brightness from the north pole only, as in
all previous panels, while panel (f) is a complete sky map containing both poles. In
calculating the sky map brightness, the synchrotron pitch angle $\theta_p$ is taken to be
zero.

First of all, the pattern of the sky map from the FF field, as seen in Figure
\ref{fig:TPCSM60_0}, is very different from the sky map from TPC model using
the vacuum magnetic field (\citealt{DyksRudak03,DHR04}; see also Fig. \ref{fig:VacSM}
below). As noted in BS10, the aberration formula used in earlier works needs to be
corrected, and this reduces the strength and increases the width of the caustics. The
FF sky map here is also very different from the corrected vacuum sky map (see Figure 4b,
4d of BS10, or Fig. \ref{fig:VacSM} below). The main differences include: 1) There is no
caustic or local enhancement produced near the NS; 2) All caustics are formed near the LC.

It is important to understand why the FF field gives such dramatically different
appearance of the sky map even though the structure of the FF field is similar to
that of the vacuum field close to the star. We find that this effect can be
attributed to the shape of the polar cap. Recall that in Figure \ref{fig:polarcap},
the FF polar cap is more circular than its vacuum counterpart. More importantly, the
FF polar cap is larger. The caustics in the vacuum field are significant in the
open field line region, but weaken toward the LOFLs, tending to disappear
if the emission is calculated from the field lines further in the closed zone.
As the larger FF polar cap encloses the vacuum polar cap, the last open field lines
of FF would map into the closed field zone of the vacuum magnetosphere. This explains
the lack of caustics close to the NS surface using the FF field.

Next we discuss the structure of the FF sky map for the TPC model (Fig. \ref{fig:TPCSM60_0}).
Each pole contributes to two main caustic structures. For the field lines coming from the
north pole (Fig. \ref{fig:TPCSM60_0}c-e), the location of the caustics roughly coincides
with the green ring, and in the sky map it appears as a broad arc (Fig. \ref{fig:TPCSM60_0}e).
At early phases ($\phi < 120^\circ$), the caustic is mainly caused by the sweepback of field
lines on the sky map (yellow field lines in Fig. \ref{fig:TPCSM60_0}a-c). This caustic is
relatively weak and narrow. At later phases, there is a bright ``knot" where emissions from
neighboring field lines cross each other on the sky map (near the intersection of light blue
field lines and the red or green rings in Fig. \ref{fig:TPCSM60_0}c). This knot and the
adjacent region contribute to the most of the second caustic, which is brighter and wider.
The knot can also be seen in panel (a), where light blue field lines fall close to each other.
There is large shear between neighboring field lines, corresponding to large current flow. The
field lines resulting in the knot enter the current sheet, and this feature persists in many
different simulations\footnote{We note that \citet{CK10} observed similar features as
they evolve the system for longer time. We've also checked simulations with still higher
resolution, and with finite conductivity using the prescription by \citet{Gruzinov07b}, and
find that the ``knot" appears in all simulations.}.

Adding the emission from both poles, there can be up to
four peaks in the gamma-ray light curve. A large fraction of the area of
the sky map includes contributions from both poles. Besides the two bright
caustics, emission from other regions is mostly smooth, except in
regions close to the LC [e.g., the tips of red and cyan lines in
panel (a)]. Comparing with Figure \ref{fig:cusheet}, we find that
these regions are deeply embedded in the strong current layer,
where the field structure may not be well resolved in the simulation.

\begin{figure}
    \centering
    \includegraphics[width=85mm]{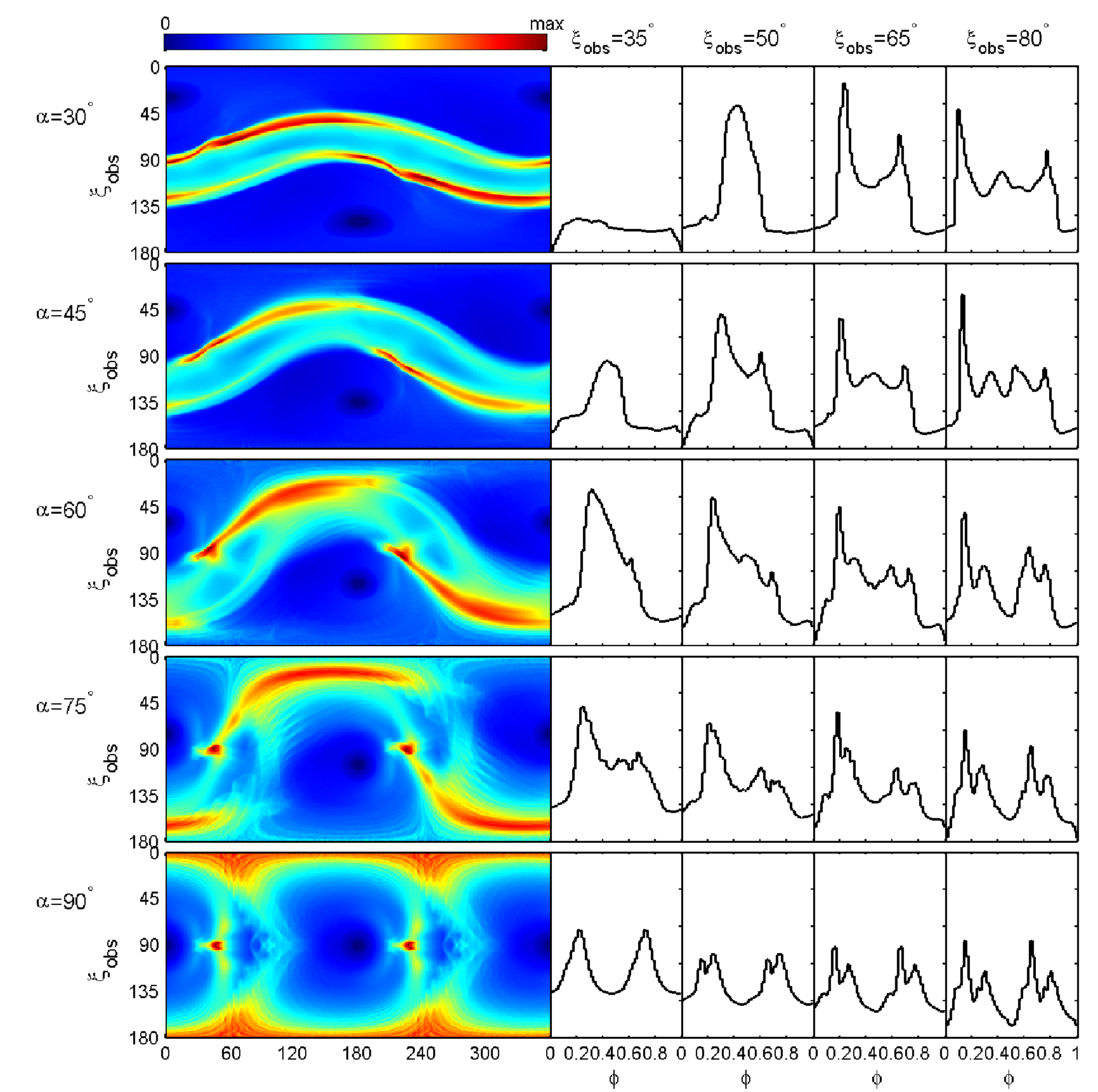}
  \caption{The light curves (right) predicted from the TPC model using the
  FF field at various magnetic inclination angles $\alpha$ and observer
  viewing angles $\zeta_{\rm obs}$. The corresponding sky maps are shown on
  the left. Emission zone is centered on LOFLs, extending to LC. A
  synchrotron pitch angle $\sin\theta_p=0.1$ is applied. The curves at the
  same $\alpha$ have the same normalization.}\label{fig:TPCLC_10}
\end{figure}

Figure \ref{fig:TPCLC_10} shows a gallery of representative light curves predicted
from the TPC model in FF field for different inclination angles and observer viewing
angles. In this plot we have chosen $\sin\theta_p=0.1$ to show the effect of non-zero
pitch angle. We see that this smoothes the sky map substantially, as can be seen by
comparing the $\alpha=60^\circ$ sky map in this figure with Figure \ref{fig:TPCSM60_0}.
Nevertheless, the light curves generally exhibit irregular shapes with up to 4 peaks.
The central part of the light curves is typically brighter because both poles contribute.
The caustic structure due to field line sweepback is smoothed and mixed with the
overlapping region from another pole, while the caustic from the vicinity of the ``knot"
is still prominent. Also note that some of the troughs in the light curves are due to
the finite size of the poles.

Comparing the light curves of Figure \ref{fig:TPCLC_10} with observations,
one generally fails to obtain a clean double-peak profile. At larger
inclination angles, the trend is to have two very broad humps instead of narrow
peaks. We have tried to restrict the geometry of the emission zone by
tuning the parameters $r_{\rm max}$ and $R_{\rm max}$, and found that they
do not improve this situation. Moreover, we have also considered simulations
with higher resolution and different diffusivity; the appearance of the sky map
and the bulk behavior of the light curves does not change qualitatively between
different runs.

\section[]{The Separatrix Layer Model}\label{sec:AG}

\begin{figure*}
    \centering
    \includegraphics[width=170mm,height=170mm]{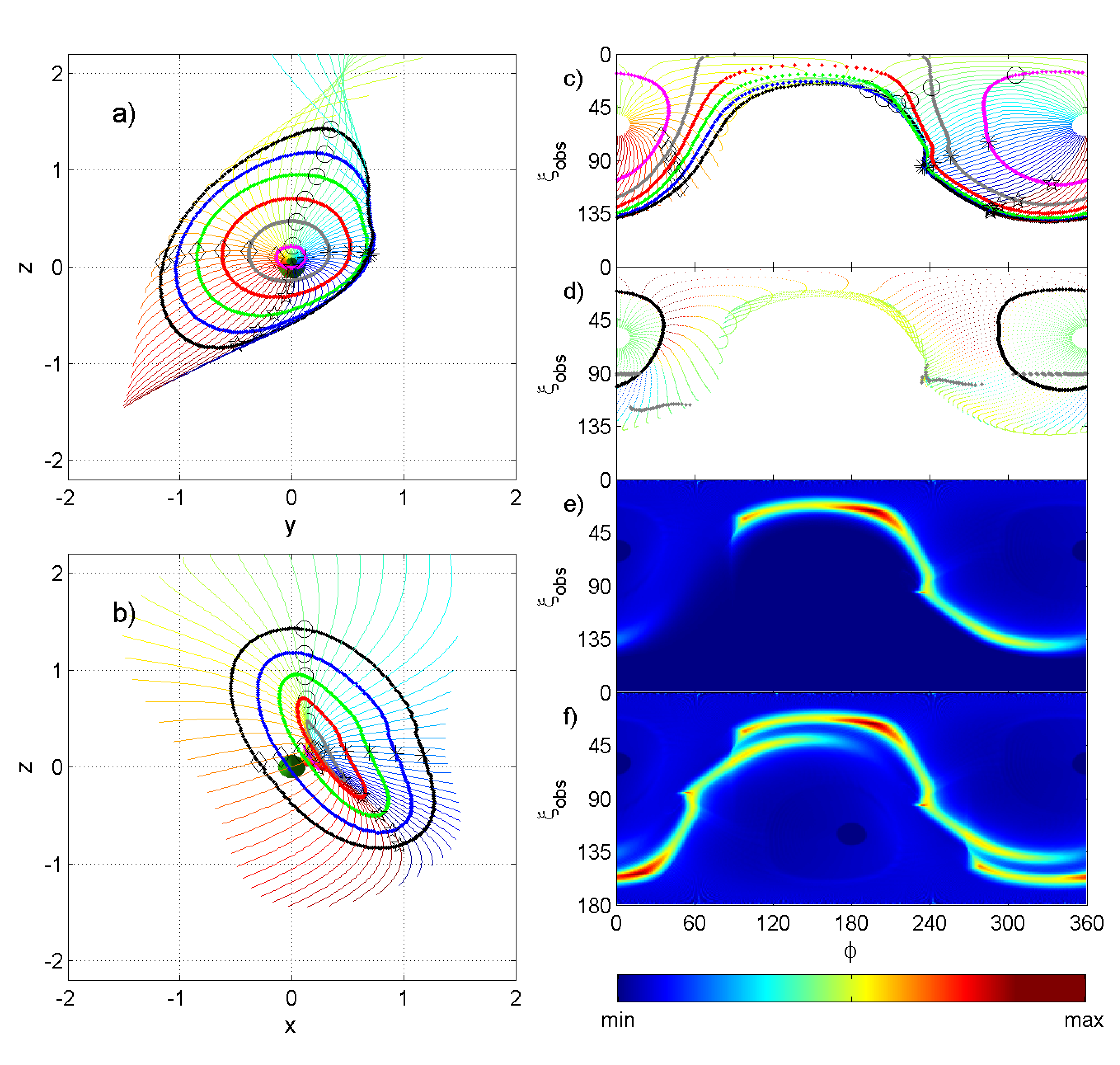}
  \caption{Same as Figure \ref{fig:TPCSM60_0}, but for the SL model, with
  inclination angle $\alpha=60^\circ$. Field lines all have $r_{\rm ov}=0.9$
  and are traced to the cylindrical radius $R=1.2R_{LC}$.}\label{fig:AGSM60_90}
\end{figure*}

The difficulties in obtaining double-peaked light curves from last open field
lines with the two-pole caustic model in FF field caused us to consider different
locations for the emission region. In this section, we study the emission from a
set of field lines in the open volume that are still close to the LOFLs. This is
motivated by the distributed nature of the return current in oblique rotators. We
will refer to this model as the ``separatrix layer" (SL), as the emission region is
concentrated in a layer in the vicinity of the separatrix. In practice, we choose
$r_{\rm ov}=0.9-0.95$ for the SL model. In Figure \ref{fig:AGSM60_90},
we plot the field lines and the corresponding sky maps with $r_{\rm ov}=0.9$,
for inclination $\alpha=60^\circ$.  The emission zone extends from the NS surface
to cylindrical cut-off radius $R_{\rm max}=1.2R_{LC}$. We organize the six panels
in the same way as in Figure \ref{fig:TPCSM60_0}.

The SL model produces two bright and narrow caustics on the sky map.
Comparing the TPC sky map with that of the SL model (i.e., moving from
$r_{\rm ov}=1.0$ to $r_{\rm ov}=0.9$), we see that the weak caustic due
to field line sweepback on the left of Figure \ref{fig:TPCSM60_0}c no
longer exists, while the caustic associated with the ``knot" has evolved to
a strong caustic spanning a large portion of observer's viewing angle. The
``knot" is no longer present. Combining the contribution from the two poles
gives rise to two caustics in panel (f).

The origin of the caustics can be traced from panels (a) to (c): we clearly
see that the red, green, blue and black rings overlap at the locations of
strong caustics. Therefore, the formation of the caustics is not due to the
coincidence that emission from {\it different} field lines congregates on the
sky map (as is the case for caustics in the vacuum field form), but the
emission from {\it one} field line arrives simultaneously, piling up on the
sky map. We call this effect ``sky map stagnation" (SMS). In Figure
\ref{fig:AGSM60_90}, we see that SMS starts to develop at a distance from the
star along field lines of $0.75R_{LC}$ (beyond the red ring). This corresponds
to the outer magnetosphere. The SMS phenomenon covers about one half ($\pi$
radians) of magnetic azimuth on each pole. Therefore, the full caustic structure
caused by SMS sweeps a large fraction of the sky map, giving rise to two sharp
peaks in the resultant light curve.

In Appendix \ref{app:stagnation} we show that SMS is a natural consequence of
the rotating split monopole solution of \citet{Michel73b}. The direction of
particle motion in this field is exactly radially outward, and the winding of
the field lines into spirals compensates for the time delay effect. The FF
field of the aligned rotator \citep{CKF99} approaches the split monopole
solution beyond the LC. The oblique rotator also asymptotes to the inclined
split monopole solution (\citealp{Bogovalov99}, S06), and this causes the
caustics in a general FF field. To check this, we have examined the field
geometry of open field lines near and beyond LC. We find that the field
resembles well the split monopole field. The deviations, which are stronger
for lines with larger $r_{\rm ov}$, prevent SMS from occurring on all open field
lines. We note, that split monopole field is just one out of many field line
configurations that can result in sky map stagnation (see Appendix
\ref{app:stagnation}).

We also constructed sky maps from different open volume coordinates, ranging
from $r_{\rm ov}=0.95$ to $r_{\rm ov}=0.60$. We find that for smaller values of
$r_{\rm ov}$ the SMS behavior is stronger. At $r_{\rm ov}=0.60$, SMS is achieved
for all field lines. This suggests that as we shift towards the center of the
open flux tube, the split monopole solution provides a better approximation of
the field geometry (see also Figure \ref{fig:rov_cmp}). We also examined the sky
map from other inclination angles ranging from $\alpha=15^\circ$ to $90^\circ$
(see also Figure \ref{fig:AGLC_10} below). We find that SMS is in fact a general
feature of the FF field.

One consequence of SMS is that the intensity of the caustics is proportional
to the length of particle trajectories within the emission zone, if emissivity
is constant along the trajectory. We have chosen to cut off the radiation at
$R_{\rm max}=1.2R_{LC}$, which is arbitrary. Choosing larger cutoff radius will
result in brighter peaks, and vice versa. As emphasized in \S\ref{sec:model}, we
have used the assumption of constant emissivity along particle trajectories rather
than field lines. The two different treatments do not produce significant
differences at radii of order of $R_{LC}$ (our approach gives slightly fainter
peaks); nevertheless, this approach is more self-consistent, and sets the
framework for future work (e.g., spectral calculations).

Another consequence of SMS is that the curvature radius $R_{\rm cur}$ of the
particle trajectories becomes much larger than the curvature radius of magnetic
field lines. One extreme example is the curvature radius of the trajectory in
split monopolar field, which equals infinity, even though the field lines are
curved. We have calculated $R_{\rm cur}$ along the open field lines of the FF
field (see Appendix \ref{app:geometry} for methodology), and found that
$R_{\rm cur}$ generally increases as particle travels farther along the
field line, and can reach about $10R_{LC}$ at the LC. Such large curvature
radius makes curvature radiation potentially inefficient, and a non-zero pitch
angle helps with radiation efficiency, which justifies our inclusion of this
effect.

We comment on the inner annular gap (IAG) model by \citet{Qiao_etal04,Qiao_etal07}.
The choice of $r_{\rm ov}=0.90$ is consistent with the IAG region. However, the
emission of the IAG model is thought to originate relatively close to the NS
surface, up to the null charge surface (NCS). As seen from Figure \ref{fig:TPCSM60_0},
no caustic can form in the inner region of the magnetosphere since the field line
is still far from approaching the asymptotic split monopole regime. The IAG model
also suffers from the more serious problem that only a small fraction of field
lines cross the NCS (see \S\ref{sec:OG}).

\begin{figure*}
    \centering
    \includegraphics[width=170mm,height=170mm]{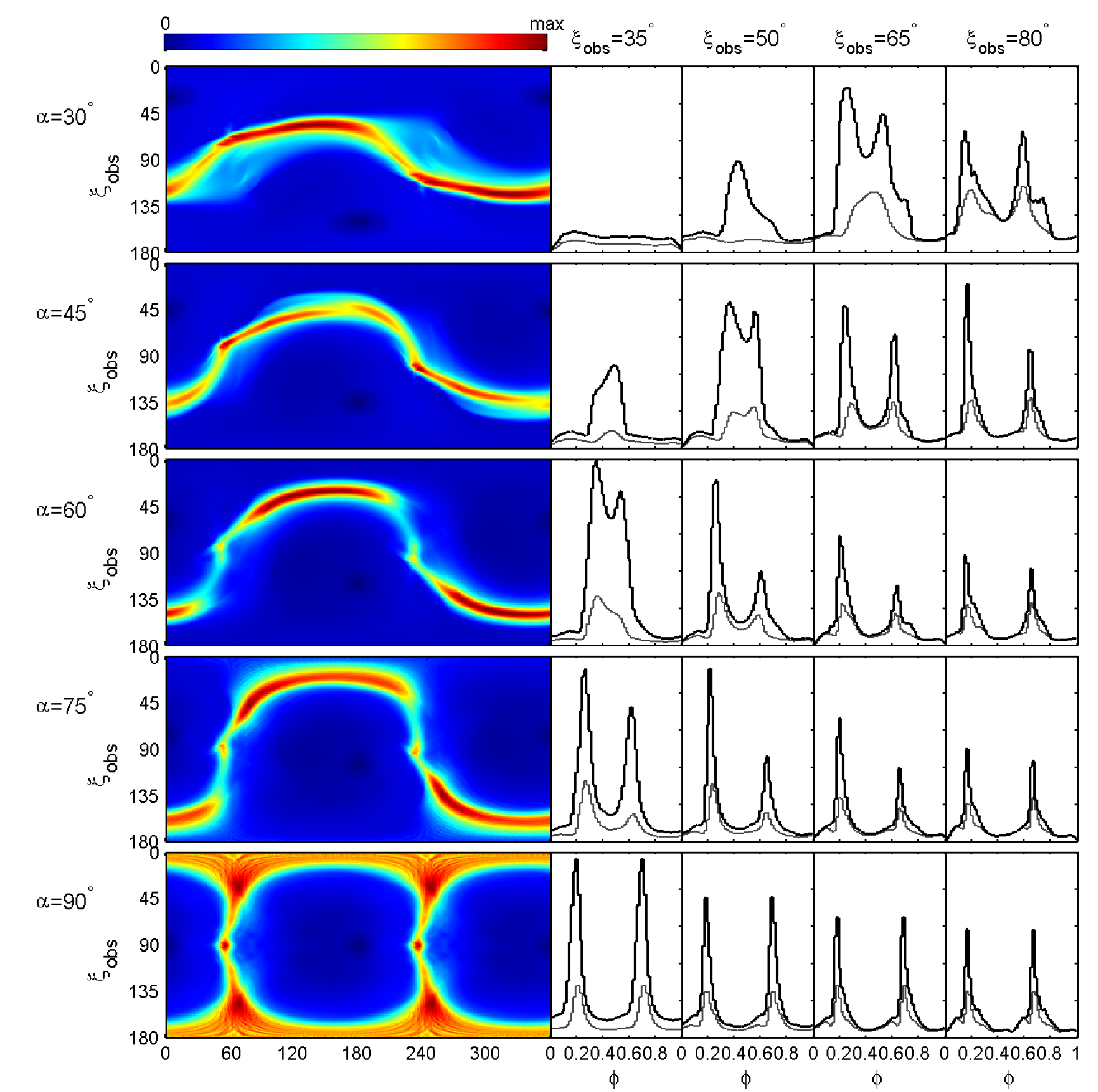}
  \caption{Same as Figure \ref{fig:TPCLC_10}, but for the SL model. For
  inclination angles $30^\circ$ and $45^\circ$, we've chosen $r_{\rm ov}=0.9$,
  and the rest use $r_{\rm ov}=0.95$. For all angles, $\sin\theta_p=0.1$. For
  the sky maps and the light curves in black, the emission is traced to
  cylindrical radius $R=1.5R_{LC}$. Grey curves show the contribution from the
  inner magnetosphere, $R<0.9R_{LC}$.}\label{fig:AGLC_10}
\end{figure*}

Figure \ref{fig:AGLC_10} collects the light curves of the SL
model for five different inclination angles and four observer
viewing angles. The emission is centered on $r_{\rm ov}=0.9$
(for $\alpha=30^\circ$ and $45^\circ$) and at $r_{\rm ov}=0.95$
(for $\alpha=60^\circ$, $75^\circ$ and $90^\circ$). The emission
is prescribed up to cylindrical radius $R_{\rm max}=1.5R_{LC}$.
Due to SMS, the intensity of the calculated light curves depends
on how far the emission is traced. To examine the relative importance
of emission contributed from different regions, we also plot the
light curves originating from within cylindrical radius
$R_{\rm max}=0.9R_{LC}$, indicated as grey lines. There are several
features worth noting as we discuss below.

First, we see that the double peak structure is a common feature for
most of the light curves. Similar to the conventional TPC and OG models
using the vacuum field (\citealt{Watters09}), the double peak feature
is most likely to appear  at relatively large $\alpha$ and $\zeta_{\rm obs}$.
Second, we see that emission outside the LC dominates the peak brightness,
while the bridge emission is from inside the LC. This is expected from SMS.
As a consequence, the intensity contrast at peak and bridge reflects how
far the separatrix layer extends beyond the LC. For many viewing angles, the
first peak tends to be stronger than the second peak. This is due to our
assumption of constant emissivity and simplified choice of emission zone
boundaries. More physics has to be included to accurately constrain the
peak intensities. Finally, without smoothing due to the non-zero pitch
angle, the SMS tends to produce reflection-asymmetric light curves. That
is, both peaks have a sharp rise and a relatively smooth decay. This appears
to be inconsistent with some of the observed gamma-ray pulsar light curves
[e.g., light curve of the Vela pulsar \citep{FermiVela}, which has a ``horn"
structure, symmetric upon reflection around the middle of the bridge]. By adding
a small pitch angle, the force-free light curves are smoothed and become more
reflection-symmetric (e.g., the light curve with $\alpha=60^\circ$ and
$\xi_{\rm obs}=50^\circ$).

\begin{figure}
    \centering
    \includegraphics[width=90mm]{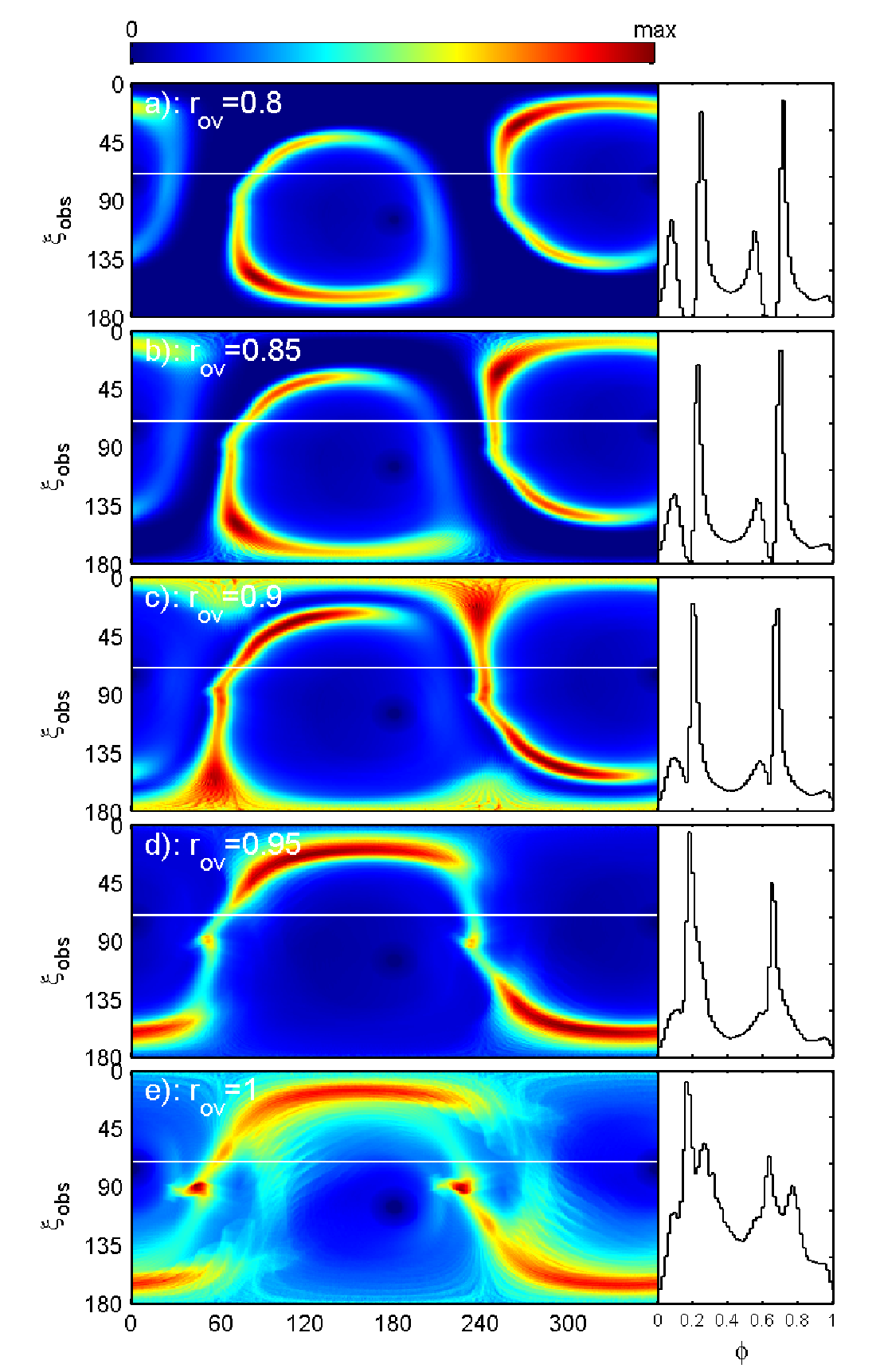}
  \caption{A comparison of sky maps with different $r_{\rm ov}$. The
  inclination angle is $\alpha=75^\circ$. Field lines are traced
  to $R=1.2R_{LC}$ except for $r_{\rm ov}=1.0$ which is traced to
  $R=1.0R_{LC}$. $\sin\theta_p=0.1$ is used for all plos. On the
  right shows the light curves with $\xi_{\rm obs}=70^\circ$.}\label{fig:rov_cmp}
\end{figure}

It remains to explain why we choose $r_{\rm ov}$ between $0.90$
and $0.95$, justifying the emission zone as being in the ``separatrix
layer". In Figure \ref{fig:rov_cmp}, we show a series of sky maps
with inclination $\alpha=75^\circ$ from $r_{\rm ov}=0.80$ to
$r_{\rm ov}=1.0$, using $\sin\theta_p=0.1$ smoothing. From the top panel,
we see that at small $r_{\rm ov}$ SMS develops on all field lines, but
only covers a relatively small fraction of the sky map. Emission from
two poles does not overlap. Consequently, 4 peaks appear in typical light
curves. As one increases $r_{\rm ov}$, on the one hand, the sky map
coverage from each pole increases, reducing the gap region between the two
poles; on the other hand, the number of field lines that exhibit SMS
decreases, reducing the intensity of two of the four peaks. At around
$r_{\rm ov}=0.90$ to $0.95$, emission from both poles starts to overlap
on the sky map, and SMS remains strong for a number of field lines. Upon
increasing $r_{\rm ov}$ further towards last open field lines, a significant
fraction of the sky map is contributed to by two poles, SMS weakens, and the
prominent ``knot" appears. A similar trend is observed for simulations with
other inclination angles.

In sum, we have shown that the separatrix layer model with $r_{\rm ov}$ between
$0.90$ and $0.95$ is capable of producing double peak light curves in a wide
range of geometric parameters. We will discuss the origin of SL emission in
\S\ref{sec:discussion}. The fact that the emission zone responsible for the
caustics in the separatrix layer scenario is located in the outer magnetosphere
is reminiscent of the outer gap model. However, there are two key differences
between the SL model and the conventional OG. First, the OG model is based on the
charge-separated magnetosphere, where the formation of the gap is thought to occur
only on field lines that cross the NCS. On the other hand, the SL model assumes that
the emission zone extends all the way from the stellar surface to beyond the light
cylinder on all field lines at a given open volume flux surface, regardless of
whether these lines cross the NCS. Second, at a fixed line of sight, most emission,
especially the two peaks, are contributed by one pole in the conventional OG model.
In contrast, in the above SL model, both poles contribute to the observed emission,
particularly, with each pole contributing one peak. This is similar to the TPC model.

\section[]{The Outer-Gap Model}\label{sec:OG}

\begin{figure}
    \centering
    \includegraphics[width=85mm]{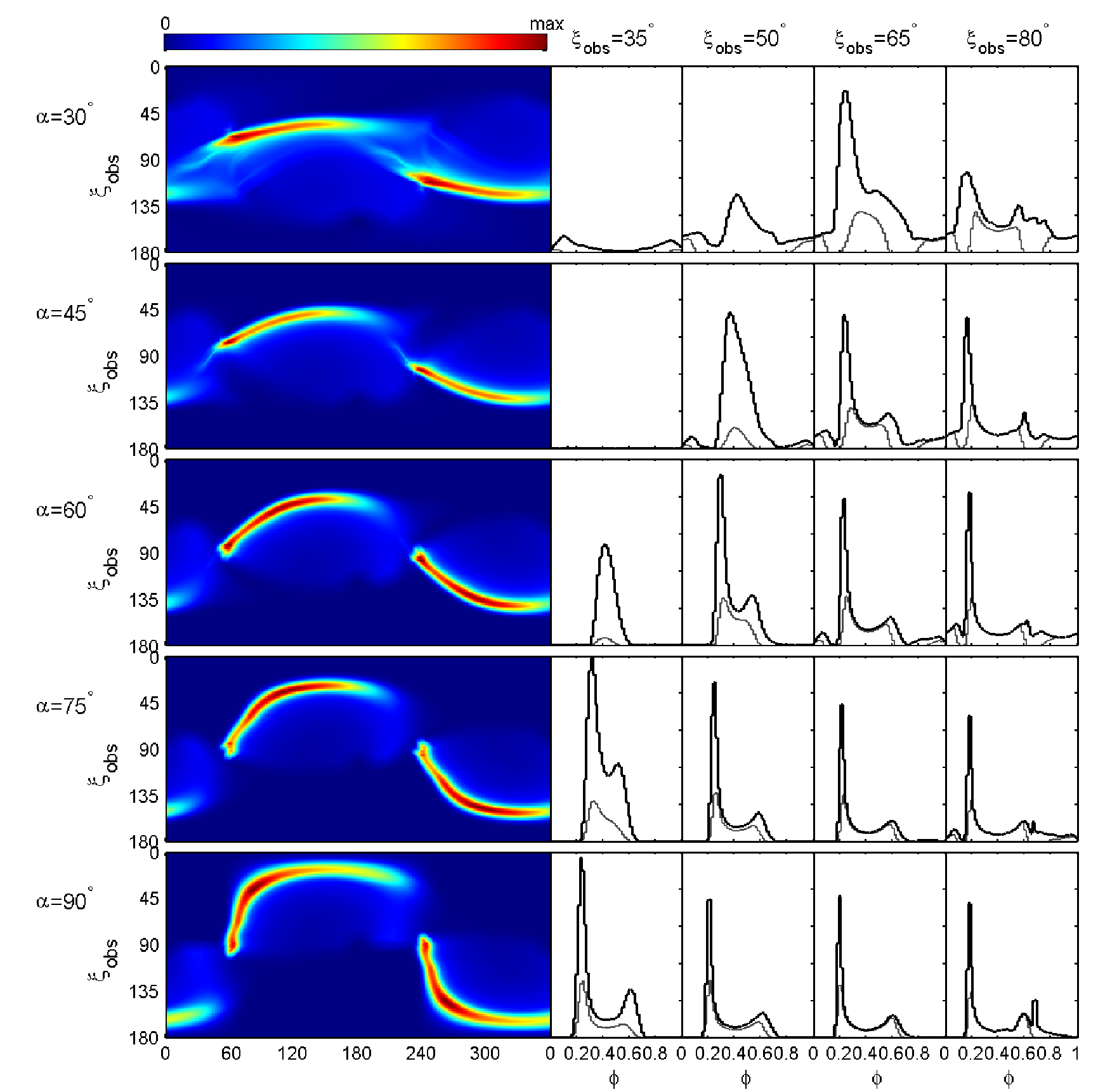}
  \caption{Same as Figure \ref{fig:AGLC_10}, but for the OG model,
  and emission zone is centered on $r_{\rm ov}=0.90$.}\label{fig:OGLC_10}
\end{figure}

In this section we discuss the sky maps and light curves from the OG
model using the FF field. A key ingredient of the OG model is the null charge
surface (NCS), which determines both the radiating field lines and the extent
of emission on these field lines. In the case of vacuum dipole field, NCS is
simply the surface where $B_z=0$. As a result, emission from the north pole
mostly contributes to the $90^\circ<\zeta_{\rm obs}<180^\circ$ portion of the
sky map, and the south pole mainly contributes to the other half. In the FF
magnetosphere, we can determine the NCS by finding where $4\pi\rho=\nabla\cdot{\mb E}=0$.
Guided by the recent work that proposed shifting the inner boundary of the OG towards
the star (e.g., \citealt{Hirotani07}), we allow emission from the full length of each
field line that crosses the NCS. This does not result in extraneous peaks as can happen
in the vacuum field, but just maximizes the sky map coverage of the OG model, improving
the bridge emission.

In Figure \ref{fig:OGLC_10}, we show the sky map of the OG model for
FF field with $r_{\rm ov}=0.90$. Because the separatrix layer model includes
emission from {\it all} field lines with this $r_{\rm ov}$, the sky map
of the OG model is simply a subset of that of the SL model. In fact,
Figure \ref{fig:OGLC_10} is just part of Figure \ref{fig:AGLC_10},
except the value of $r_{\rm ov}$ there is slightly different ($0.9$ vs $0.95$).
Assigning emission only to NCS-crossing field lines almost always eliminates
one peak from the light curve. The remaining peak is from the SMS effect of the
south pole. The north pole contributes very little to the light curves.

\begin{figure*}
    \centering
    \includegraphics[width=150mm]{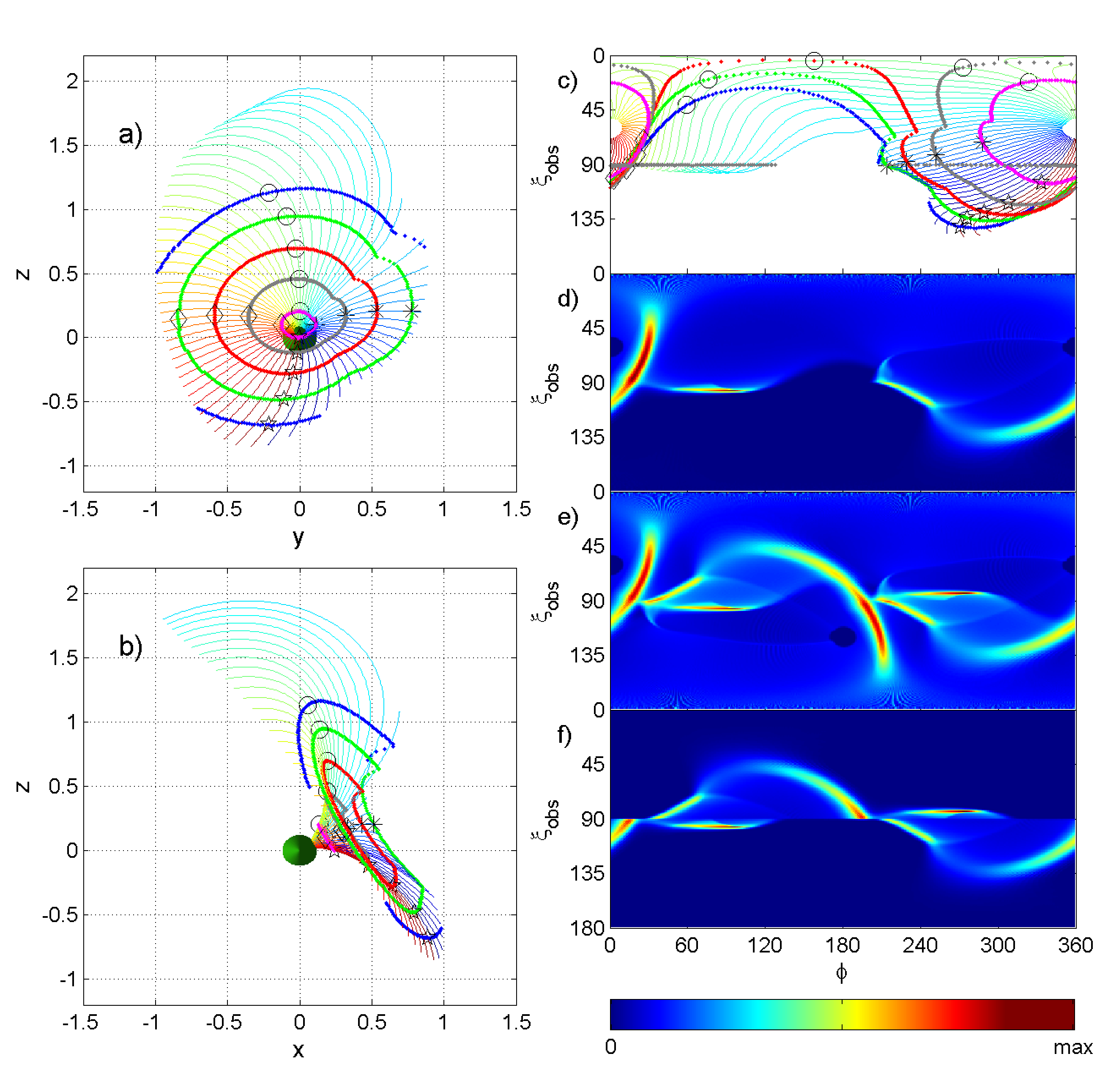}
  \caption{Sky map and field line structure using vacuum field presented for
  comparison with the SL model of the FF field from Fig. \ref{fig:AGLC_10}.
  The inclination angle is $\alpha=60^\circ$. Field lines all have
  $r_{\rm ov}=0.9$ and are traced to the LC (cylindrical radius $<R_{LC}$).
  a) and b): Field lines from the north pole viewed from two different
  directions. c): Line plot of the sky map based on field lines from the north
  pole. Line colors are same as in a,b). Color rings represent points located
  at the same distance along the magnetic field lines from the center of NS:
  pink (0.25$R_{LC}$), grey (0.50$R_{LC}$), red (0.75$R_{LC}$), green (1.0$R_{LC}$),
  blue (1.25$R_{LC}$), black (1.50$R_{LC}$). Asterisks, circles, diamonds and stars
  are added to help in identifying field lines. Gray diamonds at
  $\zeta_{\rm obs} \sim 90^\circ$ show the location of the null charge surface on
  these field lines. d): Sky map from the north pole that results from c) after
  including emission on every field line at $r_{\rm ov}=0.9$. The emission is included
  from the star to the cutoff radius. e): Sky map that results when the emission from
  both poles is superimposed. This map incorporates emission regions of both TPC and
  OG models (formally, TPC model is usually centered on $r_{\rm ov}=1$, but the picture
  is similar at $r_{\rm ov}=0.9$). While it generally results in too many peaks, this
  plot demonstrates the origin of all caustics in the vacuum field. The clustering of
  rings in panel c) indicates enhanced caustic emission in the sky map, although here
  the caustics are caused by near overlap of neighboring field lines, rather than SMS
  from each field line. f): Sky map that results from e) when emission below null
  charge surface is removed. This is the classical OG model.
}\label{fig:VacSM}
\end{figure*}

The reason for the elimination of one peak in the OG model is best
seen in Figure \ref{fig:AGSM60_90} of the previous section. In panel
(d), the location of the NCS on the sky map from the north pole is
indicated by grey dots. We can see  that only a fraction of field lines
cross the NCS. These field lines span roughly $180^\circ$ in magnetic
azimuth on the polar cap. This is to be compared with case of the vacuum
field, as in Fig. \ref{fig:VacSM}, where for similar geometry, almost all
field lines cross the ${\mb\Omega}\cdot{\mb B}=0$ surface (presumably the
NCS) eventually. For the same $r_{\rm ov}$, the FF field lines are
``straighter" than their vacuum counterparts, and a large fraction
of them do not turn horizontal inside the LC (e.g., compare lines marked
with circles in Figs. \ref{fig:AGSM60_90} and  \ref{fig:VacSM}). Therefore,
using the FF field, half of the field lines end up being non-emitting under
the framework of the OG model\footnote{Note that the charge density inside
the artificially large stellar radius of the FF simulation ($R_N^{\rm sim}$)
in Figure \ref{fig:AGSM60_90} is simply taken from the first term on the right
hand side of equation (\ref{eq:rho}); therefore, the grey dots inside the black
circle are roughly a straight line at $\xi_{\rm obs}=90^\circ$ (corresponding to
$B_z=0$). Outside the black circle, the grey dots are generally below
$\xi_{\rm obs}=90^\circ$, indicating that the location of the NCS in the FF field
is further out than determined from the classical ${\bf \Omega}\cdot{\bf B}=0$
criterion. The difference is more significant in regions marked with diamonds
than regions marked with asterisks. Although there is a jump in the location of
the NCS on the left of panel (d) in Figure \ref{fig:AGSM60_90} due to the large
size of the star in the simulation, it is also clear from the plot that it does
not affect our selection of NCS-crossing field lines.}. Consequently, half of the
bright caustics in panel (e) of Figure \ref{fig:AGSM60_90} are missing.
Combining the two poles, there is only one bright peak left. This is a
dramatic difference from the results using the vacuum field.

In Figure \ref{fig:OGLC_10} we confirm that for all inclination angles,
most of the light curves have only one prominent peak. In many of these
light curves, the shape of the peak as well as the bridge emission is
very similar to those produced by the OG model using the vacuum field,
but the second peak is missing, or degrades into a weak hump. The weak
hump is produced from the same pole as the main caustic, and mostly
due to the continuation of the main caustic (e.g., the panel with
$\alpha=60^\circ,$ $\xi_{\rm obs}=50^\circ$). Given the weakness of the
hump, it is unlikely to play an important role when more physics is
added to relax the assumption of constant emissivity.

In addition, we have experimented with the effect of inward emission.
This effect was used in the original version of the OG model
\citep{CHR86a,CHR86b}, and re-considered in \citet{TCS08}. We find
that the inward emission itself will produce two peaks whose shapes
are similar to the conventional OG model using the vacuum field. Both
peaks are relatively weak, and the bridge emission is almost half the
brightness of the peaks. More problematically, the first peak lags
the radio peak by more than $180^\circ$. Adding to this the main peak
from the outward emission, there can be three peaks in total. Therefore,
inward emission appears unlikely to play an important role.

\section[]{Discussion}\label{sec:discussion}

We have shown that using the FF field, conventional TPC and OG model no
longer work as well as in previous calculations using the vacuum dipole
field. Alternatively, we find that the separatrix layer model, where emission
comes from open field lines that lie just in the vicinity of the separatrix
(including the strong current layer inside the LC and the current sheet outside
the LC), is very promising in reproducing the prevailing double peak features of
gamma-ray pulsar light curves. This model works due to the sky map stagnation
effect, that is unique to FF field and occurs as the field becomes increasingly
split-monopolar with radius. Our calculations of the SL model in
\S\ref{sec:AG} work mainly on the geometric grounds with little physical input.
In this section we discuss the implications of these results to pulsar
gamma-ray radiation mechanism, and compare our results with observational
data.

\subsection[]{Applicability of the FF field}\label{ssec:FFapply}

The FF field provides a more realistic model for the pulsar magnetosphere
than vacuum field because it takes into account the field distortions due
to self-consistent currents caused by pulsar spin down. The basic assumption
of the force-free approximation is that pair creation is so efficient that
it fills the magnetosphere with abundant pair plasma to make it a perfect
conductor everywhere. Real pulsar magnetospheres are more complicated due
to dissipation and reconnection in the current sheet, formation of gaps, and
the presence of differential rotation \citep{Timokhin07a,Timokhin07b}.

One way of testing the robustness of the result obtained in this
paper is to consider fields from  strong-field electrodynamics (SFE)
introduced by \citet{Gruzinov07a,Gruzinov08a,Gruzinov08b}, which
generalizes the FF equations to include dissipation. In SFE, the FF
condition ${\mb E}\cdot{\mb B}=0$ is relaxed in space-like regions, where
the Ohmic dissipation is present, while time-like regions are still
non-dissipative. The dissipation and reconnection in the FF current
sheet may be better modeled with SFE. In addition, vacuum gaps in
pulsar magnetosphere may also be considered as dissipative regions,
where ${\mb E}\cdot{\mb B}\neq0$. We have implemented SFE formulation
into our FF code by replacing the FF current by the SFE current [see
equation (6) of \citet{Gruzinov07b}]. The conductivity scalar
$\sigma=\sigma(E_0,B_0)$ is taken to be constant. The natural scale of
$\sigma$ is $\Omega/c^2$, which we denote by $\sigma=1$. We have run our
simulation with different values of $\sigma$ around 1. We find that
the overall field configuration is more smooth than the FF field
configuration. Particularly, the ``knot" in the sky map from LOFLs which
penetrate the current sheet becomes less prominent. The strong current
layer inside the LC is not changed qualitatively, but the current sheet
outside the LC is smoothed appreciably. More energy is dissipated in the
magnetosphere, and the total Poynting flux decreases by  about $10\%$
over about $1R_{LC}$.

Despite the fact that SFE produces smoother magnetospheric
structure, we find that the resulting sky maps and light curves do
not differ qualitatively from the results with the FF field
presented in this paper. Therefore, we believe that our result is
robust and may also be applicable to dissipative magnetospheres with
moderate levels of dissipation.

\subsection[]{The Origin of the Separatrix Layer Emission}\label{ssec:origin}

We have shown that if pulsar high-energy emission originates from
open field line regions in the vicinity of the separatrix, extending
from the NS surface to beyond the LC, two bright caustics will form due
to sky map stagnation. Currently, there is no theory directly
pointing to the existence of the SL emission; however, based on the
geometric location of the SL, there are clues about its origin.

1. Modified SG/TPC model. We note that SMS is already significant
at $r_{\rm ov}=0.95$, which is very close to LOFLs. Also, in our
calculation, the emission zone is assumed to have a Gaussian profile with
width $\Delta r_{\rm ov}=0.025$ centered on $r_{\rm ov}$. Therefore, the
edge of the emission zone can be considered on the LOFLs. This emission
zone geometry is consistent with the SG model. Therefore, if the
thickness of the pair formation front is as large as $\Delta r_{\rm
ov}=0.025-0.05$, a modified SG model may be able to produce the desired
two peak light curves. Such SG must extend at least to the LC to produce
sharp peaks. Recent SG/TPC models consider emission from smaller $r_{\rm ov}$ 
flux tubes (e.g., \citealp{Venter_etal09}). Given that in FF field only the field lines with
$r_{\rm ov}<0.95$  are truly open and do not close through the current sheet, it is likely that 
at least geometrically the SL model can be considered as an extension of TPC and invokes
the same field lines.  The physical reason for the acceleration on these lines is unlikely to
be a slot gap, however, and is more  related to reconnection in the current sheet, as
discussed below. 

2. Modified OG model. Another clue comes from the distribution of
space-like regions in the magnetosphere. Regions of space-like 4-current,
which are likely to develop instabilities and could have accelerating
fields \citep{Gruzinov07b}, are preferentially located beyond the NCS. Similar
locations are typically invoked in OG models, although for reasons that rely on
complete charge separation in the magnetosphere. Even if OG-type model can be
justified based on the sign of 4-current, it would not be sufficient to produce
two peaks -- the emission has to come from more field lines than those crossing
the NCS. Interestingly, looking at Figure \ref{fig:AGSM60_90}d, the regions
showing the pile-up due to SMS have values of $\varrho$ near zero, or
negative. If small positive values of $\varrho$ can also cause acceleration,
this could explain the emission from the whole flux tube.

3. Association with the strong current. The presence of a strong current
layer within the LC and a current sheet beyond LC are general features
of FF field. Strong dissipation and reconnection is expected in such regions
\citep{Gruzinov07a,Lyubarsky08}, and the energy release can reach a few percent
of the spin down power, probably radiated in X-rays and gamma rays \citep{Lyubarsky96}.
Using Bogovalov's (1999) current sheet solution for the inclined split-monopole,
\citet{Kirketal02} were able to reproduce the typical double-peak light curves,
indicating that the field configuration around the current sheet beyond the LC is
also geometrically favored. Based on the same model, \citet{PetriKirk05} and
\citet{Petri09} further reasonably well reproduced the optical polarization of
the Crab pulsar and phase-resolved spectrum for the Geminga pulsar.
Our result that the SL emission zone resides in the vicinity of the strong
current layer and/or current sheet (see Figure \ref{fig:cusheet}) implies a
connection with the strong current. At higher inclinations, the return current
becomes increasingly distributed over the open flux, and it is possible that some
of the acceleration takes place on the open field lines. Also, the caustics of
emission form in the outer magnetosphere, and the acceleration on these field
lines can be affected by the proximity of the Y-point. Reconnection near the
Y-point (or ``Y-ring" in 3D) can load the open field lines with plasma, and cause
field-aligned accelerating fields, akin to auroral double-layers (\citealp{Arons08}).
Emission from the equatorial current sheet beyond the LC can also be important,
as the field lines we invoke for SL  trace close to the current sheet in the
wind zone. Modeling emission from the current sheet itself is more
complicated in FF simulation, since ideal MHD doesn't apply in the current
sheet, and the magnetic field direction abruptly changes in the unresolved
current sheet, which leads to tracing errors. This will be studied more in future
work. The facts that the bulk of the SL emission is formed at and beyond the
LC, and that emission from two poles overlaps on the sky map and forms two
peaks only close to the edge of the open flux tube (for $r_{\rm ov} > 0.9$,
Figure \ref{fig:rov_cmp}), suggests that the current sheet beyond the LC is
responsible for the large part of gamma-ray emission.

Our discussion on the origin of the SL model is highly speculative. A detailed
theory of the electrodynamics of current sheets has to be constructed  to test
these scenarios and uncover the origin of the SL emission.

\subsection[]{Comparison with Observations}\label{ssec:observation}

One year after launch, {\it Fermi} LAT has greatly expanded
the number of detected gamma-ray pulsars either by identifying previously
unresolved/unidentified sources (e.g., known radio pulsars, millisecond pulsars
(MSP), EGRET sources, supernova remnants (SNR), pulsar wind nebulae (PWN),
\citealp{FermiCTA1,Fermi09a,Fermi09b,FermiRelease09b}), or by the blind search
of LAT sources \citep{FermiRelease09a}. Together with previously known gamma-ray
pulsar data, a sample of more than 40 gamma-ray pulsars is now available
\citep{FermiPulsars09}, allowing more systematic comparison between observations
and theoretical models.

Under the framework of the SL model with FF field, the double-peak light
curves can be produced over a large range of geometric parameters, as
shown in \S\ref{sec:AG}. As a geometric model, it provides robust prediction
on the temporal location of the two peaks regardless of the input physics.
The phase lag of the first peak relative to the radio peak (which likely
coincides with the location of the polar cap on the sky map) ranges from
about $\delta=0.14$ to larger than $\delta=0.3$, and the separation between
the two peaks ranges from zero (i.e., one peak) to $\delta=0.5$. Smaller
phase lag typically corresponds to larger separation, and vice versa, as
seen from the caustic structure from Figure \ref{fig:AGLC_10}. Also, larger
inclination angles tend to result in larger peak separation. Statistically,
our model predicts that the majority of light curves should have widely-separated
double peaks, pulsars with smaller peak separation are less common, pulsars
with only one peak are possible but rare. Another consequence of the SL model
is that smaller peak separation usually indicates stronger bridge emission
between the peaks. The best example of this can be drawn from the light curve
with $\alpha=45^\circ$, $\xi_{\rm obs}=50^\circ$ in Fig. \ref{fig:AGLC_10}.

Based on the current data sample of gamma-ray pulsars, the majority of them
show widely separated double-peak pulse profiles, in good agreement
with the SL model. A small fraction shows double peaks with smaller
separation [e.g., B1716-44 \citep{AGILE09a}, J0007+7303, J1459-60, J1741-0254
\citep{FermiRelease09a}], but with relatively strong bridge emission, also in
qualitative agreement with our result\footnote{Note that in Figure 2 of
\cite{FermiRelease09a}, the first peak is placed at phase of 0.3 because most
of the new gamma-ray pulsars are radio quiet.}. A few objects show only one
peak [e.g., J0357+32 \citep{FermiRelease09a}, PSR 0437-4715, PSR 0613-0200
\citep{FermiRelease09b}, J2229+6114 \citep{AGILE09b}]. Such pulse profiles may
be obtained when the observer's line of sight tangentially cuts through the
caustics on the sky map (examples include $\alpha=30^\circ, \xi_{\rm obs}=50^\circ$
in Figure \ref{fig:AGLC_10}). The small number of such pulsars agrees with the
small chance of such tangential incidence. Noticeably, the fraction of single
peak MSPs appears to be higher than for normal gamma-ray pulsars
\citep{FermiRelease09b}. This may reflect the fact that MSPs tend to have
smaller inclination angle which increases the chance of tangential incidence.
Given the relatively old age of the MSP population, their smaller inclination
angle might be caused by alignment torques \citep{DavisGoldstein70,Goldreich70,Melatos2000}.

The distribution of phase lags from the SL model is not directly comparable
with the latest data since most new gamma-ray pulsars are radio-quiet. For
available radio-loud pulsars, we find some agreement, but not always. For
example, the phase separation between peaks in B1706-44 appears to be too
small to fit our model, as both peaks come in the first half of the rotation.
There may be an additional phase lag from the radio peak due to high altitude
radio emission from energetic pulsars \citep{WeltevredeJohnston08}. Therefore,
it is uncertain whether observed phase lag can be used as a reliable diagnostics.

\section[]{Summary and Conclusion}

This paper is concerned with high-energy emission from gamma-ray
pulsars modeled by the force-free (FF) field. The FF field takes into
account the effect of conducting plasma, and the charges and currents
are solved self-consistently using time-dependent simulations. By
construction, the FF magnetosphere does not emit because
${\mb E}\cdot{\mb B}=0$ everywhere, but it provides a more realistic
approximation to the field structure. The FF field differs substantially
from the vacuum dipole field in several ways. First, there is a larger
magnetic flux on the open field lines. As a result, the polar cap is
larger than the vacuum polar cap. Second, beyond the LC, FF field has a
strong current sheet and field lines approach inclined rotating split-monopole
shape. Inside the LC, there is a strong current layer coinciding with the last
open field lines. This layer is a thin current sheet for the aligned rotator, which
connects to the current sheet outside the LC. The amount of current carried by
the current sheet inside the LC decreases with the inclination angle. The inner
current sheet disappears completely for the orthogonal rotator.
Current distribution on the polar cap varies with $\alpha$, with larger fraction
of the current closing through the polar cap at larger inclinations, rather than
in a thin boundary layer. Third, the charge distribution in the FF magnetosphere
contains significant concentration of charge in the current sheet/strong current
layer, and the charge distribution in the current sheet for the aligned rotator is
substantially different from that of the oblique rotators. Moreover, the
null charge surface (NCS) is no longer determined by ${\mb \Omega}\cdot{\mb
B}=0$, and a large fraction of open field lines does not cross the NCS
at all. Fourth, the 4-current in substantial fraction of the FF magnetosphere
is space-like, requiring the presence of charges of both signs, which is
inconsistent with completely charge-separated picture.

The differences in the field structure of FF and vacuum fields give rise
to substantial differences in resulting theoretical light curves. The polar
cap in the FF magnetosphere is more circular and larger than the vacuum
polar cap. Therefore, close to the star, the location of LOFLs of the FF
field corresponds to closed field line regions in the vacuum field. Due
to this effect, the caustics seen in the two-pole caustic (TPC) model using
the vacuum field no longer exist in force-free field. Instead, the pattern
on the sky map is more complicated: up to four peaks (caustics) may exist,
and a large fraction of the sky map is covered by emission from both poles.
In addition, there is a bright ``knot" region on the sky map associated with
the strong current. In any case, we find it difficult to reproduce the
typical double-peak profile in the resulting light curves using the TPC model.

We have shown that if we relax the emission zone in the TPC model
into a separatrix layer (SL), where the emission comes from the open
field line regions near the strong current layer/current sheet and
extends from the NS surface up to and beyond the LC, two bright
caustics will appear. The cause of the caustics is completely
different from that in the case of a vacuum field. For a large fraction
of open field lines in the outer magnetosphere, the emission from
different heights along the field line piles up at the same spot on the
sky map, which strongly enhances the sky map brightness at that location.
This effect, which we call ``sky map stagnation" (SMS), is due to the
fact that the FF field asymptotically approaches the split monopole
field at large radii. As a result, the intensity of the caustics due to
SMS effect is determined by the extent of the emission zone. The typical
feature of the light curves from this SL model is two sharp peaks with
more or less uniform off-peak emission. Each pole contributes to one peak.
The rise of the peak is usually sharp while the drop is more smooth. Adding
the effect of non-zero pitch angle smoothes the sky map. This model
provides robust prediction on the location and the separation of the two
peaks, and the resulting light curves match the general features of most
observed gamma-ray pulsar light curves. Although there is no theory
predicting the existence of this separatrix layer emission, its location in
the pulsar magnetosphere strongly suggests its association with the strong
current layer and/or current sheet, particularly the current sheet beyond
the light cylinder.

The sky map from the OG model can be considered as part of the SL
model, restricted only to field lines that cross the NCS. We find,
however, almost always that only one peak can form. The reason is
that a substantial fraction of the open field lines do not cross
the NCS at all. The coverage of the emission on the sky map is
similar to that of vacuum field, but the formation of caustics is
due to SMS. Some gamma-ray pulsars  recently observed by {\it Fermi}
telescope show only one prominent peak, and the OG model using the
FF field may be relevant to such pulsars. The ``inner annular gap"
(IAG) model also relies on field lines that cross the NCS, and thus
suffers the same problem as the OG model. However, because the
emission zone in the IAG model is relatively close to the NS, where
SMS does not develop, it will be even more difficult to produce any
peaks with this model using FF field.

In constructing models of the emission zones, we assigned emission
to flux tubes concentric to the last open field lines. This can be
an oversimplification, and the actual emitting flux tube can have a
more complicated shape. This is particularly so if reconnection in
the Y-point region affects neighboring field lines in the outer
magnetosphere. The origin of those field lines on the polar cap does
not have to be concentric to LOFLs. An avenue that we have yet to
explore in detail is to assign emission not per field line, but in
the whole volume of the magnetosphere, weighted by some proxy of the
magnetospheric dissipation, such as the strength of the current.
The gross features of the force-free sky maps, such as the caustic
formation in the outer magnetosphere due to sky map stagnation will
likely remain the same; however, quantitative shape of the light
curves may be affected. This will be the subject of future work. In
addition, our modeling of pulsar's high-energy emission assumes constant
emissivity along particle trajectories, which works on purely geometric
basis and serves as the first step in comparing different theoretical
models and identifying possible regions of emission. More physics input
is needed for a consistent study of pulsar's high-energy emission that
would reveal the origin of pulsar magnetospheric accelerator.

\acknowledgments

We thank Jonathan Arons and Yuri Lyubarsky for helpful discussions. We
are also grateful to Ioannis Contopoulos for his constructive suggestions.
This work is supported by NASA grants NNX08AW57G and NNX09AT95G.
AS acknowledges the support of Alfred P. Sloan Foundation fellowship. XNB
acknowledges support from NASA Earth and Space Science Fellowship.

\appendix

\section{A. Derivation of Useful Relations in Force-Free Magnetosphere}\label{app:derivation}

Formal proof of equations (\ref{eq:ff1}) and (\ref{eq:ff2}) can be found
in \citet{Gruzinov06}. In this appendix we summarize Gruzinov's proof and
further explain the physical meaning of $\lambda$.

Both equations follow from two assumptions: a) The electromagnetic field
is force-free; b) The electromagnetic pattern corotates at angular velocity
$\Omega$. Condition b), when applied to an arbitrary vector field ${\mb U}$,
can be written as
\begin{equation}
[\pa_t+({\mb\Omega}\times{\mb r})\cdot\nabla]{\mb U}
={\mb\Omega}\times{\mb U}\ .\label{eq:corot}
\end{equation}
For electric and magnetic fields, after some algebra, the above equation can be
reduced to
\begin{equation}
\frac{\pa{\mb E}}{\pa t}=\nabla\times[({\mb\Omega}\times{\mb r})\times{\mb E}]
+4\pi\rho({\mb\Omega}\times{\mb r})\ ,\label{eq:corotE}
\end{equation}
\begin{equation}
\frac{\pa{\mb B}}{\pa t}=\nabla\times[({\mb\Omega}\times{\mb r})\times{\mb B}]
\ .\label{eq:corotB}
\end{equation}

In line with \citet{Gruzinov06}, equation (\ref{eq:corotB}), together with the
induction equation, implies ${\mb E}=-\vec\beta_0\times{\mb B}+\nabla\chi$, where
$\chi$ is a scalar function to be determined. The fact that the NS is a perfect
conductor ensures $\chi$ to be constant on the stellar surface, while the FF condition
requires $\chi$ to be constant along magnetic field lines. Therefore, $\chi$ must
be constant everywhere, completing the proof for equation (\ref{eq:ff1}).

\citet{Gruzinov06} proved equation (\ref{eq:ff2}) by using the variational principle.
Despite the mathematical beauty of this method, the physical meaning of $\lambda$ is
not explicitly seen. Below we prove equation (\ref{eq:ff2}) from direct calculation.

We begin by substituting equation (\ref{eq:corotE}) into the Ampere's law of Maxwell's
equation, and making use of equation (\ref{eq:ff1}), one immediately obtains
\begin{equation}
\nabla\times[{\mb B}+\vec\beta_0\times(\vec\beta_0\times{\mb B})]={\mb j}-
\rho({\mb\Omega}\times{\mb r})\ .\label{eq:lambda1}
\end{equation}
It remains to show that ${\mb B}\parallel({\mb j}-\rho({\mb\Omega}\times{\mb r}))$. This
becomes obvious since the FF current can be cast into [cf. equation (2) of \citet{Spitkovsky06}]
\begin{equation}
{\mb j}=\rho\frac{{\mb E}\times{\mb B}}{B^2}c+{\mb j}_\parallel\ ,\label{eq:current}
\end{equation}
where ${\mb j}_\parallel$ is the current density that is parallel to ${\mb B}$. Because
of equation (\ref{eq:ff1}), the corotation velocity ${\mb\Omega}\times{\mb r}$ can be
decomposed into the ${\mb E}\times{\mb B}$ drift velocity and a component that is parallel
to ${\mb B}$.

In sum, we have reached a remarkably simple and clear formula
\begin{equation}
\lambda{\mb B}={\mb j}-\rho({\mb\Omega}\times{\mb r})\equiv{\mb j}^{C}\ ,\label{eq:lambda2}
\end{equation}
where ${\mb j}^{C}$ is the current density in the corotating frame (CF)
\citep{Schiff39,Gron84}. Therefore, in CF, the magnetic field line and
current are parallel to each other. The ratio between $j^C$ and $B$ is denoted
by $\lambda$, and is constant along the magnetic field lines.

\section{B. Current Sheet in the Force-Free Field}\label{app:cusheet}

In this appendix we show that the current in the current sheet outside the LC in FF field
does not necessarily connect to the star. Our discussion adopts the \citet{Bogovalov99}'s
solution. It describes the field structure of a rotating split monopole,
and asymptotically approaches to the FF field at large radii. The Bogovalov's solution
reads
\begin{equation}
{\mb B}_m=\frac{f_0}{r^2}({\mb e}_r-R{\mb e}_\phi)\eta(r,\theta,\phi)\ ,\label{eq:monopole}
\end{equation}
where the subscript ``$m$" represents monopole, $f_0$ is an arbitrary constant, $R=r\sin\theta$
is the cylindrical radius, and
\begin{equation}
\eta(r,\theta,\phi)=D(\sin\alpha\sin\theta\sin(\phi+r/R_{\rm LC})+\cos\alpha\cos\theta)
\equiv D(x)\ ,\label{eq:eta}
\end{equation}
where $D(x)=1$ for $x>0$ and $D(x)=-1$ for $x<0$.

The current sheet lies on the surface determined by $x=0$, namely,
\begin{equation}
\cot\theta=\tan\alpha\sin(\phi+r/R_{\rm LC})\ .\label{eq:theta}
\end{equation}
Clearly, we have $\pi/2-\alpha\leq\theta\leq\pi/2+\alpha$. One can find the normal direction of this
surface to be
\begin{equation}
{\mb n}=\frac{1}{\sqrt{1+2A^2}}(A{\mb e}_r-{\mb e}_\theta+A{\mb e}_\phi)\ ,\label{eq:normal}
\end{equation}
where
\begin{equation}
A=[1+\tan^2\alpha\sin^2(\phi+r/R_{\rm LC})]\tan\alpha\cos(\phi+r/R_{\rm LC})\ .\label{eq:Afac}
\end{equation}

The natural frame to calculate the current in the current sheet is the CF, where the current is
given by equation (\ref{eq:lambda1}). In addition to the curl of ${\mb B}$, the CF current has
contribution from the curl of $\vec\beta_0\times{\mb E}$ term. Therefore, it is useful to define
\begin{equation}
\begin{split}
{\mb H}_m\equiv&{\mb B}_m+\vec\beta_0\times(\vec\beta_0\times{\mb B}_m)\\
=&\frac{f_0}{r^2}[(1-R^2){\mb e}_r-R{\mb e}_\phi]\eta(r,\theta,\phi)\ .\label{eq:fulcurrent}
\end{split}
\end{equation}
The surface current density in the current sheet in the CF is then given by
\begin{equation}
{\mb J}={\mb n}\times({\mb H}_m^+-{\mb H}_m^-)\ ,
\end{equation}
where superscripts ``$+$" and ``$-$" denote the quantities just above and beneath the current
sheet.

It is useful to look at two special cases, namely, the aligned and orthogonal rotators. For
$\alpha=0$,  we have ${\mb n}={\mb e}_z$. From the above equations, it is obvious that ${\mb J}$
has both radial and azimuthal components. The radial component connects the current to
the star, and decreases as $1/r$. The azimuthal component asymptotically approaches a
constant. Integrating over a circle with radius $r$, one obtains the net current carried by
the current sheet that connects to the star
\begin{equation}
I_0=\int_0^{2\pi}J_r\bigg|_{\alpha=0}rd\phi=4\pi f_0\ .\label{eq:I0}
\end{equation}
Note that $I_0$ is a constant that does not depend on $r$, reflecting the conservation of current.
One can also check that $I_0$ equals to the amount of current flowing from the star (outside the
current sheet), but with a minus sign, as expected:
\begin{equation}
I_{{\rm star},0}=\oint_{\theta\neq\pi/2}(\nabla\times{\mb H})\bigg|_{\alpha=0}
\cdot {\mb e}_rr^2\sin\theta d\theta d\phi=-I_0\ .
\end{equation}

For $\alpha=90^\circ$, $A\rightarrow\infty$, thus ${\mb n}=({\mb e}_r+{\mb e}_\phi)/\sqrt{2}$.
Plugging into the above equations, one finds that ${\mb J}$ has only the $\hat{\theta}$
component. This means that the current in the current sheet is not connected to the star,
but forms poloidal loops in the $\hat{\theta}$ direction. In the FF field from our simulations,
we do observe such loops in the current sheet outside the LC. Since they are not
connected to the star, it becomes clear that there is no current sheet in the inner
magnetosphere for the orthogonal rotator.

\begin{figure}
    \centering
    \includegraphics[width=70mm,height=60mm]{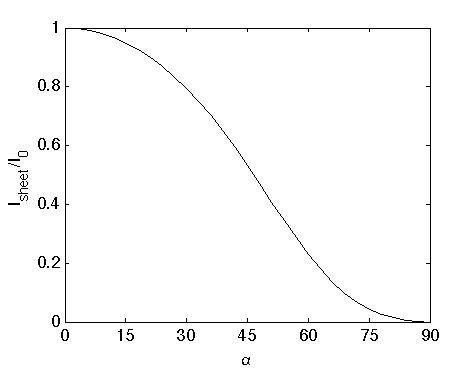}
  \caption{The amount of current in the current sheet of the rotating split monopole that
  is connected to the star $I_{\rm sheet}$ as a function of magnetic inclination angle
  $\alpha$. $I_{\rm sheet}$ is calculated from equation (\ref{eq:Isheet}), normalized to
  the value for the aligned rotator $I_0$, given by equation (\ref{eq:I0}).}\label{fig:curfrac}
\end{figure}

After some algebra, we obtain the general expression of the current carried by the current
sheet that is connected to the star
\begin{equation}
\frac{I_{\rm sheet}(\alpha)}{I_0}=\int_0^{2\pi}\frac{\sin\theta}{\sqrt{1+2A^2}}\frac{d\phi}{2\pi}\ ,
\label{eq:Isheet}
\end{equation}
where $\theta$ is determined by equation (\ref{eq:theta}) and $A$ is given by equation
(\ref{eq:Afac}). The quantity does not depend on $r$, as expected. Moreover, one can check
that this quantity also equals to the net current flowing from the the rest of the star
\begin{equation}
\frac{I_{\rm star}(\alpha)}{I_0}=-\frac{1}{2\pi}\oint\cos\theta\sin\theta D(x)d\theta d\phi\ ,
\end{equation}
where $x$ is given by equation (\ref{eq:eta}). The dependence of $I_{\rm sheet}$ on
inclination angle $\alpha$ is shown in Figure \ref{fig:curfrac}. We see that the amount of
current in the current sheet that is connected to star monotonically decreases with $\alpha$.
This is related to the degradation of the current sheet inside the LC discussed in \S\ref{ssec:rhoj}.

\section{C. Segment Length and Curvature Radius of Particle Trajectories}\label{app:geometry}

Here we describe the method for calculating the length and curvature
radius of the particle trajectories. For simplicity, we consider
particles with zero pitch angle. The discussion here is valid so far
as particle pitch angle is small, as we use in this paper. The
direction of particle motion consists of a combination of
${\mb E}\times{\mb B}$ drift velocity (or corotation velocity) and
velocity parallel to the magnetic field
\begin{equation}
{\mb e}=\vec\beta_d+\alpha{\mb t}=\vec\beta_0+\alpha'{\mb t}\ ,\label{eq:a1}
\end{equation}
where ${\mb t}\equiv\pm{\mb B}/B$ is the unit vector parallel to the magnetic
field line (without loss of generality, hereafter we assume the radial component
of ${\mb t}$ points outward), and $\alpha$ or $\alpha'$ are obtained by requiring
$|{\mb e}|=1$:
\begin{equation}
\alpha=-\vec\beta_d\cdot{\mb t}\pm\sqrt{(\vec\beta_d\cdot{\mb t})^2+1-\beta_d^2}\ ,\qquad
\alpha'=-\vec\beta_0\cdot{\mb t}\pm\sqrt{(\vec\beta_0\cdot{\mb t})^2+1-\vec\beta_0^2}\ ,
\label{eq:a2}
\end{equation}
where the plus/minus signs correspond to trajectories along/opposite to magnetic
field line. Usually, we expect $\vec\beta_0\cdot{\mb t}<0$ due to field sweepback.
Therefore, the outgoing particles always have $\alpha'>0$, and are expected to
have $\alpha'>1$ (for reference, split monopole field has $\alpha'=\sqrt{1+(R/R_{LC})^2}$).
For ingoing particles, however, $\alpha'$ is positive inside the LC,
negative outside the LC. This means that ingoing particles can return to the star
{\it only} if they are generated inside the LC. Below we consider only the outgoing
particles, taking the plus sign in the equations above.

Particle trajectories can be obtained by tracing the direction of motion while
rotating the star at the same time. Since ${\mb e}$ has a corotation component,
equations (\ref{eq:a1}) ensure that particle trajectories follow the field lines
exactly in the corotating frame (CF). Although this is obvious, we still go
through the proof, which will be useful for numerical calculations of curvature
radius of particle trajectories.

\begin{figure}
    \centering
    \includegraphics[width=50mm,height=45mm]{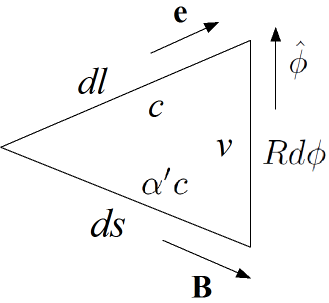}
  \caption{Graphic illustration for particle trajectory calculation.}\label{fig:geometry}
\end{figure}

The proof is best demonstrated in a graphical view. In equation (\ref{eq:a1}),
the vector equation forms a triangle as shown in Figure \ref{fig:geometry}.
The length of each edge is expressed in terms of velocity. Multiplying each
of the edges by an amount of time $dt$, we obtain another interpretation of the
graph. The upper edge gives an infinitesimal segment length $dl=cdt$ along
particle trajectory in the direction of ${\mb e}$. The lower edge becomes
$ds=\alpha'cdt$, which marks the corresponding segment length along the
magnetic field line. As the particle travels by $dl$, the star (and the field
lines) rotate by an angle $d\phi=\Omega dt=\Omega dl/c=dl/R_{LC}$. Therefore,
the entire field line is shifted by $Rd\phi=\beta_0cdt=Rdl/R_{LC}$ in the
azimuthal direction, this corresponds to the right edge. Now we've obtained a
closed triangle that is similar to the original velocity triangle. Based on
this new interpretation, we conclude that particle trajectory follows the
magnetic field line exactly in the CF.

The fact that particle trajectories follow CF magnetic field lines as well as the
similarity relations from the velocity and segment triangles provide a natural way
to reconstruct particle trajectory. The key relation is between $dl$ and $ds$,
which satisfies
\begin{equation}
dl=ds/\alpha'\ .\label{eq:a3}
\end{equation}
In practice, we originally have a series of segment lengths along field lines $s[i]$.
From this formula, one can obtain a series of segment lengths of the corresponding
particle trajectory $l[i]$. This further provides the phase of rotation as the
particle travels, $\phi[i]$. Finally, particle trajectory can be reconstructed by
rotating each point on the magnetic field line by $\phi[i]$. It is then
straightforward to (numerically) calculate the curvature radius of the particle
trajectory.

When performing the calculation, however, there can be cases where the square root
in equation (\ref{eq:a2}) becomes imaginary. This always happens in the vacuum
dipole field once the particle travels out of the LC; it also happens occasionally
in the current sheet since the ideal MHD doesn't apply there and one does not expect
equation (\ref{eq:a1}) to hold. Moreover, across the FF current sheet, the toroidal
component of the magnetic field changes sign, resulting in ambiguity in defining
``ingoing" and ``outgoing" directions. In such regions, the proposed ``outgoing"
particles may have $\alpha'<1$, which is not desired. Therefore, we modify equation
(\ref{eq:a3}) into
\begin{equation}
dl=ds/\max(\alpha',1)\ .\label{eq:a4}
\end{equation}
In calculating $\alpha'$, if the terms under the square root is negative, we just
take them to be zero.

One can further show the analytical expression for the curvature radius of particle
trajectory based on the discussions above:
\begin{equation}
\frac{1}{R_c}=\bigg|\frac{{\mb \Omega}\times{\mb e}}{c}+\alpha'\frac{d{\mb e}}{ds}\bigg|\ .
\end{equation}
Numerical experiments show that curvature radius obtained from this approach has
very similar accuracy as the first method.

\section{D. Conditions for Sky Map Stagnation}\label{app:stagnation}

According to the formula of aberration (\ref{eq:a1}) and
time delay (\ref{eq:tdelay}), the condition for emission along a
magnetic field line to become stagnant on the sky map is
\begin{equation}
{\mb B}\cdot\nabla e_z=0,\qquad {\mb B}\cdot\nabla(\phi_e+{\mb
r}\cdot{\mb e})=0\ ,\label{eq:stagnant}
\end{equation}
where $\phi_\eta$ denotes the azimuthal angle of the emission
direction $\hat\eta$.

The second equation can be reduced to
\begin{equation}
{\mb B}\cdot\nabla({\mb r}\cdot{\mb e})+[{\mb e}\times({\mb
B}\cdot\nabla){\mb e}]_z/(1-e_z^2)=0\ .\label{eq:stag2}
\end{equation}

Note that the emission direction depends only on the direction of
the magnetic field, but not its strength. Therefore, the only variable
left is the direction of the field, which has two degrees
of freedom. The above are two independent equations for ${\mb B}$.
Therefore, the direction of ${\mb B}$ can be uniquely determined
from equations (\ref{eq:stagnant}), if initial condition is given.
Nevertheless, the equations above are still too complicated to solve
analytically. Instead of solving the equations directly, we test
certain field geometries to see if the stagnation condition is
satisfied.

We consider the split monopole field (\ref{eq:monopole}) and show that
it does indeed satisfy equations (\ref{eq:stagnant}).
According to equation (\ref{eq:a1}), the emission direction from split
monopole is exactly along the radial direction ${\mb e}^m={\mb e}_r$.
Since the strength $f_0/r^2$ and sign $\eta(r,\theta,\phi)$ have nothing
to do with the emission direction ${\mb e}^m$, we can safely omit it in
the calculations below.

Now we substitute the split monopole field into these conditions.
The first equation in (\ref{eq:stagnant}) becomes ${\mb
B}^m\cdot\nabla\cos\theta=0$. Note that $B^m_{\theta}=0$; therefore,
this condition is satisfied. For equation (\ref{eq:stag2}), the
first term is just $B_r=1$, while the second term can be shown to be
equal to $-R\sin\theta/r(1-\sin^2\theta)=-1$, and they sum to zero.
Therefore, emission from any {\it one} field line in the split monopole
field is projected exactly to the same point on the sky map.

Obviously, the split monopole is not the only possible solution to
equations (\ref{eq:stagnant}). It only corresponds to a special initial
condition. In fact, we see in Figure \ref{fig:AGSM60_90} that at
$r_{\rm ov}=0.9$, SMS starts to develop from inside the LC, where the
FF field still deviates from the split monopole solution. Nevertheless,
SMS is much easier to achieve at smaller $r_{\rm ov}$ (see Figure
\ref{fig:rov_cmp}), or at larger radii ($R\gg R_{\rm LC}$). The fact
that the FF solution approaches the inclined split monopole asymptote
\citep{Bogovalov99}, as well as the observation of SMS strongly suggest
its connection to the split monopole field geometry at large radii.

\label{lastpage}

\begin{thebibliography}{67}
\expandafter\ifx\csname natexlab\endcsname\relax\def\natexlab#1{#1}\fi

\bibitem[{{Abdo} {et~al.}(2008)}]{FermiCTA1}
{Abdo}, A.~A. {et~al.} 2008, Science, 322, 1218

\bibitem[{{Abdo} {et~al.}(2009{\natexlab{a}})}]{Fermi09a}
---. 2009{\natexlab{a}}, \apjl, 695, L72

\bibitem[{{Abdo} {et~al.}(2009{\natexlab{b}})}]{FermiVela}
---. 2009{\natexlab{b}}, \apj, 696, 1084

\bibitem[{{Abdo} {et~al.}(2009{\natexlab{c}})}]{Fermi09b}
---. 2009{\natexlab{c}}, \apjl, 699, L102

\bibitem[{{Abdo} {et~al.}(2009{\natexlab{d}})}]{Fermi09c}
---. 2009{\natexlab{d}}, \apj, 699, 1171

\bibitem[{{Abdo} {et~al.}(2009{\natexlab{e}})}]{Fermi09d}
---. 2009{\natexlab{e}}, \apj, 700, 1059

\bibitem[{{Abdo} {et~al.}(2009{\natexlab{f}})}]{FermiRelease09a}
---. 2009{\natexlab{f}}, Science, 325, 840

\bibitem[{{Abdo} {et~al.}(2009{\natexlab{g}})}]{FermiRelease09b}
---. 2009{\natexlab{g}}, Science, 325, 848

\bibitem[{{Abdo} {et~al.}(2009{\natexlab{h}})}]{FermiPulsars09}
---. 2009{\natexlab{h}}, arxiv:0910.1608

\bibitem[{{Arons}(1983)}]{Arons83}
{Arons}, J. 1983, \apj, 266, 215

\bibitem[{{Arons}(2008)}]{Arons08}
{Arons},J. 2008, arXiv:0708.1050

\bibitem[{{Arons} \& {Scharlemann}(1979)}]{AS79}
{Arons}, J. \& {Scharlemann}, E.~T. 1979, \apj, 231, 854

\bibitem[{{Bai} \& {Spitkovsky}(2010)}]{BS09a}
{Bai}, X.-N. \& {Spitkovsky}, A. 2010, ApJ, accepted, arXiv:0910.5740

\bibitem[{{Bogovalov}(1999)}]{Bogovalov99}
{Bogovalov}, S.~V. 1999, \aap, 349, 1017

\bibitem[{{Cheng} {et~al.}(1986{\natexlab{a}}){Cheng}, {Ho}, \&
  {Ruderman}}]{CHR86a}
{Cheng}, K.~S., {Ho}, C., \& {Ruderman}, M. 1986{\natexlab{a}}, \apj, 300, 500

\bibitem[{{Cheng} {et~al.}(1986{\natexlab{b}}){Cheng}, {Ho}, \&
  {Ruderman}}]{CHR86b}
---. 1986{\natexlab{b}}, \apj, 300, 522

\bibitem[{{Cheng} {et~al.}(2000){Cheng}, {Ruderman}, \& {Zhang}}]{CRZ00}
{Cheng}, K.~S., {Ruderman}, M., \& {Zhang}, L. 2000, \apj, 537, 964

\bibitem[Contopoulos \& Kalapotharakos(2009)]{CK10}
Contopoulos, I., \& Kalapotharakos, C.\ 2009, arXiv:0912.2369

\bibitem[{{Contopoulos} {et~al.}(1999){Contopoulos}, {Kazanas}, \&
  {Fendt}}]{CKF99}
{Contopoulos}, I., {Kazanas}, D., \& {Fendt}, C. 1999, \apj, 511, 351

\bibitem[{{Daugherty} \& {Harding}(1982)}]{DH82}
{Daugherty}, J.~K. \& {Harding}, A.~K. 1982, \apj, 252, 337

\bibitem[{{Daugherty} \& {Harding}(1996)}]{DH96}
---. 1996, \apj, 458, 278

\bibitem[{{Davis} \& {Goldstein}(1970)}]{DavisGoldstein70}
{Davis}, L. \& {Goldstein}, M. 1970, \apjl, 159, L81

\bibitem[{{Deutsch}(1955)}]{Deutsch55}
{Deutsch}, A.~J. 1955, Annales d'Astrophysique, 18, 1

\bibitem[{{Dyks} {et~al.}(2004){Dyks}, {Harding}, \& {Rudak}}]{DHR04}
{Dyks}, J., {Harding}, A.~K., \& {Rudak}, B. 2004, \apj, 606, 1125

\bibitem[{{Dyks} \& {Rudak}(2003)}]{DyksRudak03}
{Dyks}, J. \& {Rudak}, B. 2003, \apj, 598, 1201

\bibitem[{{Goldreich}(1970)}]{Goldreich70}
{Goldreich}, P. 1970, \apjl, 160, L11

\bibitem[{{Goldreich} \& {Julian}(1969)}]{GJ69}
{Goldreich}, P. \& {Julian}, W.~H. 1969, \apj, 157, 869

\bibitem[Gr{\o}n(1984)]{Gron84} Gr{\o}n, {\O}.\ 1984,
International Journal of Theoretical Physics, 23, 441

\bibitem[{{Gruzinov}(2005)}]{Gruzinov05}
{Gruzinov}, A. 2005, Physical Review Letters, 94, 021101

\bibitem[{{Gruzinov}(2006)}]{Gruzinov06}
---. 2006, \apjl, 647, L119

\bibitem[{{Gruzinov}(2007{\natexlab{a}})}]{Gruzinov07a}
---. 2007{\natexlab{a}}, \apjl, 667, L69

\bibitem[{{Gruzinov}(2007{\natexlab{b}})}]{Gruzinov07b}
---. 2007{\natexlab{b}}, ArXiv e-prints

\bibitem[{{Gruzinov}(2008{\natexlab{a}})}]{Gruzinov08a}
---. 2008{\natexlab{a}}, Journal of Cosmology and Astro-Particle Physics, 11, 2

\bibitem[{{Gruzinov}(2008{\natexlab{b}})}]{Gruzinov08b}
---. 2008{\natexlab{b}}, ArXiv e-prints

\bibitem[Gurevich et al.(1993)]{Gurevich_etal93} Gurevich, A., Beskin, V., \& Istomin, Y.\ 1993,
Physics of the Pulsar Magnetosphere, by Alexandr Gurevich and Vassily Beskin and Yakov
Istomin, pp.~432.~ISBN 0521417465.~Cambridge, UK: Cambridge University Press, August
1993.,

\bibitem[{{Halpern} {et~al.}(2008){Halpern}, {Camilo}, {Giuliani}, {Gotthelf},
  {McLaughlin}, {Mukherjee}, {Pellizzoni}, {Ransom}, {Roberts}, \&
  {Tavani}}]{Halpern_etal08}
{Halpern}, J.~P., {Camilo}, F., {Giuliani}, A., {Gotthelf}, E.~V.,
  {McLaughlin}, M.~A., {Mukherjee}, R., {Pellizzoni}, A., {Ransom}, S.~M.,
  {Roberts}, M.~S.~E., \& {Tavani}, M. 2008, \apjl, 688, L33

\bibitem[{{Harding} {et~al.}(2008){Harding}, {Stern}, {Dyks}, \&
  {Frackowiak}}]{Harding_etal08}
{Harding}, A.~K., {Stern}, J.~V., {Dyks}, J., \& {Frackowiak}, M. 2008, \apj,
  680, 1378

\bibitem[{{Harding} {et~al.}(1978){Harding}, {Tademaru}, \&
  {Esposito}}]{Harding_etal78}
{Harding}, A.~K., {Tademaru}, E., \& {Esposito}, L.~W. 1978, \apj, 225, 226

\bibitem[{{Hirotani}(2007)}]{Hirotani07}
{Hirotani}, K. 2007, \apj, 662, 1173

\bibitem[{{Kalapotharakos} \& {Contopoulos}(2009)}]{KC09}
{Kalapotharakos}, C. \& {Contopoulos}, I. 2009, \aap, 496, 495

\bibitem[{{Kirk et~al.}(2002)}]{Kirketal02} {Kirk}, J.~G. and {Skj{\ae}raasen}, O. and {Gallant}, Y.~A. 2002, \aap, 388L, 29

\bibitem[{{Komissarov}(2006)}]{Komissarov06}
{Komissarov}, S.~S. 2006, \mnras, 367, 19

\bibitem[{{Lyubarskii}(1996)}]{Lyubarsky96}
{Lyubarskii}, Y.~E. 1996, \aap, 311, 172

\bibitem[{{Lyubarsky}(2008)}]{Lyubarsky08}
{Lyubarsky}, Y. 2008, in American Institute of Physics Conference Series, Vol.
  983, 40 Years of Pulsars: Millisecond Pulsars, Magnetars and More, ed.
  C.~{Bassa}, Z.~{Wang}, A.~{Cumming}, \& V.~M. {Kaspi}, 29--37
  
 \bibitem[Lyubarsky(2009)]{Lyubarsky09} Lyubarsky, Y.\ 2009, \apj, 
696, 320

\bibitem[{{Lyutikov} \& {Thompson}(2005)}]{LyutikovThompson05ApJ}
{Lyutikov}, M. \& {Thompson}, C. 2005, \apj, 634, 1223

\bibitem[{{McKinney}(2006)}]{McKinney06}
{McKinney}, J.~C. 2006, \mnras, 368, L30

\bibitem[{{Melatos}(2000)}]{Melatos2000}
{Melatos}, A. 2000, \mnras, 313, 217

\bibitem[{{Michel}(1973{\natexlab{a}})}]{Michel73a}
{Michel}, F.~C. 1973{\natexlab{a}}, \apj, 180, 207

\bibitem[{{Michel}(1973{\natexlab{b}})}]{Michel73b}
---. 1973{\natexlab{b}}, \apjl, 180, L133

\bibitem[{{Muslimov} \& {Harding}(2003)}]{MH03}
{Muslimov}, A.~G. \& {Harding}, A.~K. 2003, \apj, 588, 430

\bibitem[{{Muslimov} \& {Harding}(2004)}]{MH04a}
 2004, \apj, 606, 1143

\bibitem[{{Pellizzoni} {et~al.}(2009{\natexlab{a}})}]{AGILE09a}
{Pellizzoni}, A. {et~al.} 2009{\natexlab{b}}, \apj, 691, 1618

\bibitem[{{Pellizzoni} {et~al.}(2009{\natexlab{b}})}]{AGILE09b}
---. 2009{\natexlab{a}}, \apjl, 695, L115

\bibitem[{{P{\'e}tri}(2009)}]{Petri09}
{P{\'e}tri}, J. 2009, \aap, 503, 13

\bibitem[P{\'e}tri \& Kirk(2005)]{PetriKirk05} P{\'e}tri, J.,
\& Kirk, J.~G.\ 2005, \apjl, 627, L37

\bibitem[{{Qiao} {et~al.}(2004){Qiao}, {Lee}, {Wang}, {Xu}, \&
  {Han}}]{Qiao_etal04}
{Qiao}, G.~J., {Lee}, K.~J., {Wang}, H.~G., {Xu}, R.~X., \& {Han}, J.~L. 2004,
  \apjl, 606, L49

\bibitem[{{Qiao} {et~al.}(2007){Qiao}, {Lee}, {Zhang}, {Wang}, \&
  {Xu}}]{Qiao_etal07}
{Qiao}, G.-J., {Lee}, K.-J., {Zhang}, B., {Wang}, H.-G., \& {Xu}, R.-X. 2007,
  Chinese Journal of Astronomy and Astrophysics, 7, 496

\bibitem[{{Rafikov} \& {Goldreich}(2005)}]{RafikovGoldreich05}
{Rafikov}, R.~R. \& {Goldreich}, P. 2005, \apj, 631, 488

\bibitem[{{Romani} \& {Yadigaroglu}(1995)}]{RY95}
{Romani}, R.~W. \& {Yadigaroglu}, I.-A. 1995, \apj, 438, 314

\bibitem[Schiff(1939)]{Schiff39} Schiff, L.~I.\ 1939,
Proceedings of the National Academy of Science, 25, 391

\bibitem[{{Spitkovsky}(2006)}]{Spitkovsky06}
{Spitkovsky}, A. 2006, \apjl, 648, L51

\bibitem[{{Spitkovsky}(2008)}]{Spitkovsky08}
{Spitkovsky}, A. 2008, in American Institute of Physics Conference Series, Vol.
  983, 40 Years of Pulsars: Millisecond Pulsars, Magnetars and More, ed.
  C.~{Bassa}, Z.~{Wang}, A.~{Cumming}, \& V.~M. {Kaspi}, 20--28

\bibitem[{{Takata} \& {Chang}(2007)}]{TakataChang07}
{Takata}, J. \& {Chang}, H.-K. 2007, \apj, 670, 677

\bibitem[{{Takata} {et~al.}(2007){Takata}, {Chang}, \& {Cheng}}]{TCC07}
{Takata}, J., {Chang}, H.-K., \& {Cheng}, K.~S. 2007, \apj, 656, 1044

\bibitem[{{Takata} {et~al.}(2008){Takata}, {Chang}, \& {Shibata}}]{TCS08}
{Takata}, J., {Chang}, H.-K., \& {Shibata}, S. 2008, \mnras, 386, 748

\bibitem[{{Tang} {et~al.}(2008){Tang}, {Takata}, {Jia}, \&
  {Cheng}}]{Tang_etal08}
{Tang}, A.~P.~S., {Takata}, J., {Jia}, J.~J., \& {Cheng}, K.~S. 2008, \apj,
  676, 562

\bibitem[{{Thompson}(2004)}]{Thompson04}
{Thompson}, D.~J. 2004, in Astrophysics and Space Science Library, Vol. 304,
  Cosmic Gamma-Ray Sources, ed. K.~S. {Cheng} \& G.~E. {Romero}, 149--+

\bibitem[{{Timokhin}(2006)}]{Timokhin06}
{Timokhin}, A.~N. 2006, \mnras, 368, 1055

\bibitem[{{Timokhin}(2007{\natexlab{a}})}]{Timokhin07a}
---. 2007{\natexlab{a}}, \apss, 308, 575

\bibitem[{{Timokhin}(2007{\natexlab{b}})}]{Timokhin07b}
---. 2007{\natexlab{b}}, \mnras, 379, 605

\bibitem[Venter et al.(2009)]{Venter_etal09} Venter, C., Harding, 
A.~K., \& Guillemot, L.\ 2009, \apj, 707, 800

\bibitem[{{Watters} {et~al.}(2009){Watters}, {Romani}, {Weltevrede}, \&
  {Johnston}}]{Watters09}
{Watters}, K.~P., {Romani}, R.~W., {Weltevrede}, P., \& {Johnston}, S. 2009,
  \apj, 695, 1289

\bibitem[{{Weltevrede} \& {Johnston}(2008)}]{WeltevredeJohnston08}
{Weltevrede}, P. \& {Johnston}, S. 2008, \mnras, 391, 1210

\bibitem[{{Yadigaroglu}(1997)}]{Yadi97}
{Yadigaroglu}, I.-A.~G. 1997, PhD thesis, Stanford University

\end{thebibliography}
\end{document}